# Avoiding the "Great Filter": An Assessment of Climate Change Solutions and Combinations for Effective Implementation

Junze Zhang[1], Kerry Zhang[2], Mary Zhang[3], Jonathan H. Jiang[4], Philip E. Rosen[5], Kristen A. Fahy[4]

[1.] Bonita High school, La Verne, CA 91750, USA
[2.] University High School, Irvine, CA 92612, USA
[3.] Diamond Bar High School, Diamond Bar, CA 91765, USA
[4.] Jet Propulsion Laboratory, California Institute of Technology, Pasadena, CA 91108, USA
[5.] Independent researcher, Vancouver, WA 98662, USA

Correspondence: Jonathan.H.Jiang@jpl.nasa.gov

## Abstract

Climate change is the long-term shift in global weather patterns, largely caused by anthropogenic activity of greenhouse gas emissions. Global climate temperatures have unmistakably risen and naturally occurring climate variability alone cannot account for this trend. Human activities are estimated to have caused about 1.0 °C of global warming above the pre-industrial baseline and if left unchecked, will continue to drastically damage the Earth and its inhabitants. Globally, natural disasters and subsequent economic losses have become increasingly impactful because of climate change. Both wildlife ecosystems and human habitats have been negatively impacted, from rising sea levels to alarming frequency of severe weather events around the world. Attempts towards alleviating the effects of global warming have often been at odds and remain divided among a multitude of strategies, reducing the overall effectiveness of these efforts. It is evident that collaborative action is required for avoiding the most severe consequences of climate change. This paper evaluates the main strategies (industrial/energy, political, economic, agricultural, atmospheric, geological, coastal, and social) towards both mitigating and adapting to climate change. As well, it provides an optimal combination of seven solutions which can be implemented simultaneously, working in tandem to limit and otherwise accommodate the harmful effects of climate change. Previous legislation and deployment techniques are also discussed as guides for future endeavors.

## 1. Introduction

The Great Filter was a concept first proposed by Robin Hanson in 1996 in an online essay titled "The Great Filter - Are We Almost Past It?". It was a theorized answer to the famous Fermi Paradox [1], essentially stating that the reason Earth has not contacted extraterrestrial civilization yet is the existence of a particular barrier that prevented most if not all life from developing to the higher levels on the Kardashev scale [2]. Certainly, there are an almost infinite number of possibilities of what this filter could be, but as the problem of global warming comes into view, many fear that the consequences of climate change may be what ended other life and will soon fall upon Earth as well.

In 1988, global warming first became a major political concern when watershed events placed the issue in the spotlight [3]. However, since impacts decades into the next century was viewed as still very far off, the concern was dismissed and the attention of the public quickly returned to more immediate matters. However, with evidence of a relationship between human actions and global warming becoming apparent, climate change gradually surfaced as a relevant political topic. Such fear was justified: potential hazards resulting from global warming included droughts, floods, hurricanes, severe storms, heatwaves, wildfires, cold spells, and landslides. Constant threats such as temperature shifts, precipitation variability, changing seasonal patterns, changes in disease distribution, desertification, ocean-related impacts, and soil and coastal degradation contribute to vulnerability across multiple sectors in many countries [4]. It was later reported that the failure to reduce emissions from such hazards will cost the world at least $2 billion per day in economic losses. In fact, the 2018 wildfires alone cost approximately equal to the collective losses from all wildfires incurred over the past decade [5].

Climate change was first officially recognized by the political world in 1989 during the Geneva climate conference, where the Intergovernmental Panel on Climate Change (IPCC) was established under the U.N. to review further political and economic consequences of global warming. Since then, milestones for individual countries have slowly been set and edited through following conferences. Though multiple world leaders claimed their countries have been putting in their best efforts to reach their stated goals, objective data shows that many such efforts were, however, far from sufficient. On November 4, 2016, the Paris Agreement went into force as an internationally recognized treaty. The Agreement requires all states to align their efforts with their "nationally determined contributions" (NDCs) that they set themselves and to report regularly on these efforts [6]. Yet 5 years later, as pointed out by Sir Robert Watson, former chair of the IPCC, most countries still need to triple their 2030 reduction commitments to be aligned with their own Paris Agreement targets. An analysis of the pledges for the 184 signatory countries found that almost 75% were insufficient [5]. Furthermore, individual solutions, although recommended, were not specified. For instance, Pakistan has not included a single measurable target in its contribution to a UN climate deal. With political drama and limited technology, governments around the globe are still debating which solutions to implement and how much time, effort, and money should be put into each. This also applies similarly to the U.S. itself when the Supreme Court on June 30th, 2022 (West Virginia v. EPA) limited the Environmental Protection Agency's ability to regulate carbon emissions from power plants, making emission reduction more difficult.

Numerous solutions have been proposed, and some complement each other better if implemented together as opposed to utilizing a single solution or a different combination. The purpose of this paper is to evaluate all aspects of these solutions and to devise the optimal combinations that are the most cost-effective, easiest to implement and would benefit humanity the most from the devastating impact of global climate variability. Integrating different solutions that complement each other limits the consequences of others while boosting their combined benefits simultaneously. For instance, a carbon tax proposal by Professor Gilbert E. Metcalf of Tufts University not only included a straightforward tax of $15 per metric ton of $CO_2$, but he also proposed three different forms of tax credits that would benefit factory workers to compensate for the tax. By this innovative approach, manufacturers would not be incentivized to increase the market price level of their goods, eliminating the major concern of raising taxes leading to higher prices. The revenue from the carbon tax – estimated to be $90.1 billion – could be reinvested in other programs [7]. Another pair of complementary solutions includes industrial enhancements in plants along with additional funding and research to maximize the efficiency of industrial capital in production. As an example, large scale stationary sources of SOx & NOx dramatically reduced emissions by installing SCRs (selective catalytic reduction) equipment on furnace stacks in the 1990s, which resulted in a landmark victory in combating smog, particulates, and acid rain [8].

Recently, the international community has categorized climate change solutions into two broad areas: mitigation and adaptation. Mitigation focuses on limiting the quantities of greenhouse gasses (GHGs) that foster climate change from being emitted into the atmosphere and decreasing the concentrations of existing gasses from the atmosphere. This is done by either regulating emission sources or enhancing Negative Emissions Technologies (NETs) [9]. Adaptation strategies stress the importance of preparing to cope with any potential hazards and evading damage. These include managing increasingly extreme conditions, protecting coastlines from rising sea levels, managing ecosystems, dealing with reduced water availability, developing resilient crop varieties, and protecting infrastructure [10]. A blend of mitigative and adaptive solutions are needed to address the risks of climate change; mitigation solutions lessen the efforts needed for adaptation, and in turn, adaptation decreases the target intensity of mitigation, thus stressing again the value of integrating multiple policies. Suggested combinations in this paper will include both adaptation and mitigation strategies.

This paper first contains a thorough evaluation of 23 commonly discussed solutions. The solutions are separated into eight categories based on their fields of technologies. The efficiency of each solution is deduced, standardized, and then used for comparison. The efficiency of a solution is usually determined by the estimated metric tons of GHGs reduced over a controlled variable of resources (capital, time, physical efforts, funding) for mitigation solutions while adaptation solutions are measured from the risk reductions in disasters (number of lives, amount of property, stability lost). In addition, the assessment also considers the solutions' unique advantages and disadvantages. Finally, the current viability and technological readiness of the strategy will also be given to show when a solution should be implemented to achieve its maximum potential. For better comparison, compiling tables are constructed to weigh the data as well. Based on the evaluations, the paper suggests an ideal combination of the assessed strategies for implementation at the international level.

## 2. Mitigation and Adaptation Solutions

### Energy

*Nuclear:*

Nuclear energy and renewable energy are currently the two pillars of clean energy. While the use of renewable energy is steadily climbing, the future of nuclear energy is much less optimistic, contributing 18% of global energy in 1996 and only 11% today. This decline in energy production is a result of increasing competition with renewable energy and low natural gas prices [11]. In addition, fear of nuclear accidents and the possibility of countries transforming nuclear power plants to develop nuclear weapons also limits the support and funding for nuclear energy.

The main advantage of nuclear energy production is a nuclear reactor's high-capacity (i.e., operational) factor. In 2021, this value was estimated to be around 92%, suggesting that each nuclear unit produces energy approximately 92% of the time on average. This generates nearly 800 billion kWh of energy in the United States annually, avoiding more than 470 million tonnes of carbon emissions each year. Nuclear energy production will also lead to substantial economic benefits; some estimate that current nuclear units located in the United States generate approximately $40-50 billion each year, providing a wide range of stable job opportunities. Furthermore, land requirements for nuclear energy production are notably lower than other clean-energy sources.

As stated in Table 1, while nuclear plants are relatively simple and inexpensive to maintain over long periods of time, capital costs for nuclear power plants are extremely expensive, ranging from $6,500-$12,250 per kilowatt for a 2,200 MW plant, and levelized cost of energy (LCOE) ranging from $112-$189 per MWh generated [12]. Simply, the high up-front cost of building each nuclear unit deters many investors and companies. In addition, false associations of nuclear power plants with nuclear weapons, as well as to past nuclear accidents, also contribute to the lack of funding and support for this type of energy resource. A fair depiction of nuclear power plants to the public is vital for widespread support, as with many other mitigation strategies.

Nuclear energy is one of the few energy sources that can provide extensive amounts of energy without damaging the environment. Media misinterpretation of scientific investigations have led to a decline in nuclear energy support, a deficient supply chain, and insufficient workforce. Reversal of this reality requires improved education and the development of a larger market.

Table 1: Nuclear Energy Data Analysis

| Source | LCOE[a] | Cost ($ per kWH)[b] | Energy required | Land required |
|---|---|---|---|---|
| Nuclear [13] | $148 (+20% from 2009) | Capital costs: $6,500 - $12,250 LCOE: $112 - $189 | 0.1 - 0.3 kWh of energy input required | Around 1 square mile for a 1,000-megawatt facility |

a: Levelized cost of electricity (total cost of building and operating over an assumed lifetime)
b: Kilowatt hours (3,600 kilojoules)

*Renewable:*

Renewable energy (RE) can be defined as originating from sources naturally restored or regenerated and regarded as zero, low or neutral in GHG emission during energy production. This energy type has been receiving the most positive attention as the use of fossil fuels is being challenged, encouraging the growth of renewable energy with support from a developing market and leading RE to become the fastest-growing energy source globally. The future of RE depends heavily on both international and domestic policies and goals. Market conditions, such as resource availability, cost, demand, and regulations, also determine the growth of renewable energies.

The main sources of RE (in order from greatest to least generation) are hydroelectric, biomass/biofuels, geothermal, wind, and solar. Each type of renewable energy encompasses its distinct set of benefits and disadvantages, and it is difficult to list out a detailed description of all commonalities between the types. However, it can be said that all of these renewable energy types generally increase job opportunities and efficiently use secure energy sources. The costs and land requirements for each of the renewable energy types are listed in the Table 2, but the land requirement and cost for wind energy is notably low. In addition, wind turbines are easy to maintain and can be sold off for fixed prices over long periods of time, enabling a steady income. However, geographical limitations and wind availability cause wind energy to be less effective among the renewable energy types.

Renewable energies have much lower energy capacity factors compared to nuclear power or fossil fuels. Some renewable energy sources are also largely intermittent; wind turbines and solar panels cannot produce electricity in the absence of the necessary ambient conditions. While battery storage of wind and solar-derived power has been proposed as one way to mitigate the effects of these limitations on consumers, this option adds an additional layer of thermodynamic inefficiency while increasing capital cost. As indicated in Table 2, below, effective installation of wind turbines and solar panels requires large amounts of land (e.g., 43.6kWH for wind turbines and 100kWH for solar panels in accordance with land use) and may distress local populations. Situating wind farms away from cities would significantly lower its cost and reduce adverse consequences. However, transmission lines must then be built to deliver wind energy to population centers, further driving up cost and reducing efficiency from line losses. Moreover, wind energy development may not be the most profitable use of land and would need to compete with other high value uses.

Historically, biomass/biofuels regularly demand substantial amounts of energy to operate, and unsustainable bioenergy practices could eventually lead to deforestation and damage natural habitats of various wildlife. Bioenergy utilization also requires considerable space, as companies need to situate production plants close to sources of biomass to reduce feedstock transportation costs. Additionally, biomass companies should be encouraged to use agricultural waste instead of growing separate organic matter to reduce their land footprint. Hydroelectric energy production is similarly restricted in establishment areas due to its necessity to be located near bodies of flowing water. Without careful planning, this can disrupt the natural flows of rivers, animal migration paths, increase water toxicity, and generally displaced both humans and animals from their local environments. While hydroelectricity can be produced for relatively low costs decades after construction, as shown in Table 2, the initial financial investment remains significantly large and large-scale construction of hydroelectric plant costs may steadily increase as land areas for reservoirs are declining. Local environments must be suited for long-term energy production and precipitation trends must be favorable for hydroelectric facilities to function properly and effectively. Much of the easily accessible locations for building hydroelectric facilities have already been developed, leaving few new opportunities for additional plants.

Table 2: Renewable Energy Data Analysis

| [13] | LCOE[a] | Cost ($ per kWH)[b] | Land required (square feet per kWH) |
|---|---|---|---|
| **Hydroelectric** | -- | $0.04 | 13,700 |
| **Biomass/biofuels** | $85 | $0.09 | 152 |
| **Geothermal** | $97 | $0.04 | 196 |
| **Wind** | $45 (-67% from 2009) | $0.07 | 43.6 (direct land use only) |
| **Solar** | $50 (-86% from 2009) | $0.10 | 100 |

a: Levelized cost of electricity (total cost of building and operating over an assumed lifetime)
b: Kilowatt hour (3,600 kilojoules)

Currently, Iceland is completely independent of fossil fuels and other nonrenewable energy sources, producing electricity only from hydropower and geothermal facilities, specifically generating 75% and 25% of its total electricity consumption, respectively. In addition, Iceland has taken advantage of domestic volcanic activity and geothermal energy to obtain hot water and heat. Interestingly, Iceland only shifted to renewable energy because of economic reasons as opposed to environmental concerns - the country could not continue to sustain expensive oil importation prices and required a stable energy source. This transition suggests that reprioritized economic policies could be significantly more effective compared to prolonged discussions about global warming consequences, placing an emphasis on political and economic solutions that favor carbon taxation and discourage pollution. Another lesson that can be obtained from this example is to utilize regional environmental advantages, such as Iceland's abundance of naturally occurring geothermal energy. For instance, Rock Port, Missouri exploits its wind resources to produce 125% of the town's energy consumption, and unused energy can then be sold to other areas as a source of income. While some areas are more suitable for the installation of solar panels, others may instead be incentivized to build geothermal plants due to local characteristics. Thus, each region should be responsible for procuring the maximum benefit based on their own natural atmospheric and geological advantages.

**Economic/Political**

*Carbon Tax:*

Under a carbon tax, the government sets a price that GHG emitters must pay for each standardized quantity of greenhouse gasses they expel to the atmosphere. Sweden, Finland, and the Netherlands have already adopted such taxes. The desired result is that businesses will take steps to reduce their emissions to avoid paying the tax. Ultimately, the goal is to design a carbon tax to best internalize the effects of emissions and to adjust the income or payroll tax for any distributive effects.

The lack of consensus can be viewed as the largest negative aspect of carbon tax as a solution. Setting the exact price of the tax often poses a considerable challenge for politicians as an equilibrium is hard to find in a dynamic economic environment due to concerns for raising consumer price levels. For example, tax proposals such as H.R. 2069, introduced by former U.S. Representative Pete Stark in 2007 to Congress, included taxes of $15.00/ton on coal, $3.25/barrel on oil, and $7.30/t on natural gas, but it eventually failed from a lack of agreement. International actions have been largely set back as well. If the tax rate is high enough to significantly reduce emissions, few, if any, countries will allow an international agency to collect the taxes. If the tax rate is low enough to make an international agency operational, however, it is unlikely to discourage significant cuts in fossil fuel usage [14]. Even academic papers abroad do not provide a consensus view on the marginal damages of GHG emissions and the optimal tax rate for the US either. For instance, the IPCC reports that $12 per ton would be sufficient, Stern Review reports that at least $85 is needed to implement an efficient carbon tax, while MIT researchers proposed an $18 solution with an increase of 4% per year [15].

If implemented, however, carbon taxes provide arguably the most economic return of any solution. For instance, one estimation predicts that a carbon tax starting at $25 per ton and rising at 2% over inflation annually would have raised $1 trillion over its first decade [16]. The U.S. currently raises a similar amount with all its other excise taxes. Projections of another study state that a carbon tax levied on all energy-related carbon emissions at a rate of $50 per metric ton and an annual growth rate of 5 percent would generate $1.87 trillion in federal revenue over the next 10 years [17]. Furthermore, the same study also estimates that $CO_2$ emissions will reduce by 8.4% while total greenhouse gasses would be reduced by 14%. Reductions in coal consumption would be 59%, petroleum 34%, and natural gas by 8%. And this benefit will continue; simulations show that the carbon tax revenue for the United States stays constant after 3 or 4 decades at around 1.2% of the US GDP, which is equivalent to roughly 300 billion dollars currently [7].

However, these calculations are based on participation only by the U.S. The impact of a carbon tax on the entire world depends on the number of countries agreeing to implement the tax, their tax levels, and the ways revenues would be used. For example, rebating the revenues directly back to households in poor economic conditions, using them to aid and improve the welfare in low-income communities, or compensating workers in carbon-intensive industries are some applications of the carbon tax revenue by the government [18]. Carbon tax revenues should be used to reduce other taxes in a way that maintains progressivity. A universal tax level is not recommended. However, we would propose an international system of carbon taxes that sets the taxation level for an individual country relative to its economic conditions.

Many other solutions have potential to complement a carbon tax. In fact, comprehensive carbon policy packages have already been proposed by various professionals and credited sources. Increased spending on energy-related research and development, providing energy production subsidies that contribute to a continuing reliance on US fossil fuels, and implementing tax credits are all evidence-backed suggestions for complementary implementation [19].

*Cap & Trade:*

Cap-and-trade is a term that represents a system of solutions which include an implemented "cap", or limit, on GHG emissions while simultaneously encouraging actions of "trade", or exchange, of quantities of emissions between producers. The cap represents the ceiling in which individual firms and factories are allowed to emit their greenhouse gasses. The limit would decrease over time and companies who exceed this limit would be financially penalized. The cap is thus relatively rigid. The trading aspect, however, accommodates this and makes the overall system uniquely flexible. It essentially allows firms to buy and sell the government caps with one another, meaning companies could conduct trade in their own favor to either profit or avoid additional expense.

California began operating a cap-and-trade program in 2013. The program was one of the first in the world and is among the largest. California's greenhouse gas production has fallen 5.3% between 2013 and 2017, as California targets economy-wide carbon neutrality by 2045. This improvement has not come at the expense of industrial output, with the state's manufacturing production increasing from $250 billion to $299 billion over the same period. To accommodate the recent downturn in global energy demand due to the pandemic, the state reduced the need for allowances as facilities have been producing fewer emissions.

The main benefit of a cap-and-trade system is its flexibility because it incentivizes businesses to choose the most cost-effective ways to stay under the cap while keeping compliance costs low. As shown in Figure 1, we can conduct a hypothetical scenario of two companies, Firm A and Firm B. They exist such that Firm A emits x tons of GHG per year while Firm B emits y tons. Due to their status as carbon-emitting producers, the firms have had "caps" imposed upon them by the government, thus constituting their GHG emission allowances. In this scenario, Firm A receives an allowance of x-50 tons, meaning Firm A is over-emitting by 50 tons; Firm B receives y+50 tons, meaning it has 50 tons of allowance left for utilization. Further, we assume the potential fine Firm A will receive is $100,000 for exceeding its cap, but Firm B does not have to pay a fine due to it being well below its cap. Using the "trade" system, the firms can form a pact that will result in mutual benefit. Firm A could purchase 50 tons of allowance from Firm B for $50,000 so that Firm A's emissions would not exceed its new "cap" and would not have to pay the $100,000 fine, resulting in a net expenditure avoidance of $50,000. Firm B would also benefit from the direct sale of its unused extra allowance, resulting in a revenue gain of $50,000. Both firms not only keep their emissions in their respective caps but also profit from this exchange simultaneously. In comparison to the carbon tax, under cap-and-trade businesses can seek to mitigate their emission fines whereas a carbon tax mandates a rigid, set rate for every ton of $CO_2$ emitted.

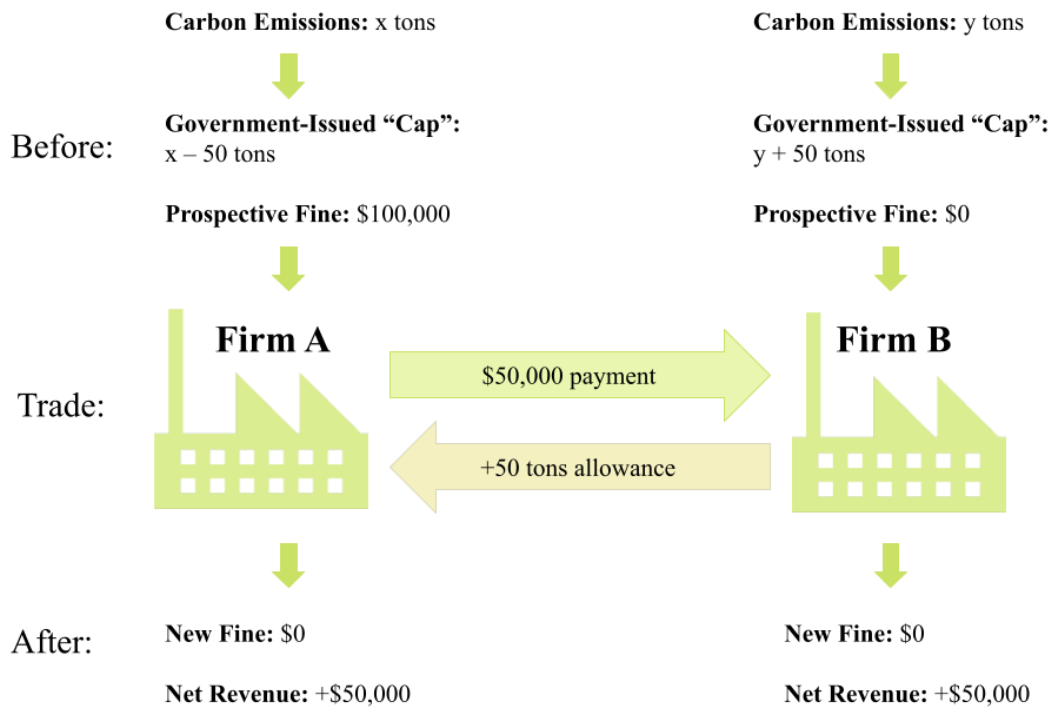

**Figure 1** - Demonstrated hypothetical diagram of emission allowances between industry firms. This shows the trading system of carbon allowances and how they may mitigate fines via the cap-and-trade system.

The cap-and-trade system also comes with its own drawbacks. When the government forces a complementary solution like renewable energy to the businesses along with cap-and-trade instead of letting businesses choose for their own interests, the market will react negatively. Additionally, a systematic market approach may fail, preventing emitters subject to a cap-and-trade system from choosing the lowest-cost compliance options. As Ann Carlson explains, if no market failure exists, policymakers should recognize the trade-off inherent in limiting the market mechanisms [20]. According to Richard Schmalensee of MIT and Robert N. Stavins of Harvard University, "…in several systems, the ability to bank allowances for later use has been an important source of cost savings. The ability to bank provides a margin of intertemporal flexibility with positive economic and environmental consequences. Changes in economic conditions can render caps non-binding or drive prices to intolerable levels." [21].

*Organizational & Private Investments:*

In August 2007, Secretariat of the United Nations Framework Convention on Climate Change published a technical paper, *Investment and Financial Flows to Address Climate Change*, estimating that approximately $205 billion dollars in additional investment will be required annually by 2030 to meet emissions reduction targets. There exist previous efforts in climate investing such as the World Bank Group, the world's largest contributor to climate investment for developing countries, with $26 billion in 2021. The Climate Investment Funds (CIF), which includes $8.5 billion dollars in total, describes its goal "to accelerate climate action by empowering transformations in clean technology, energy access, climate resilience, and sustainable forests in developing and middle-income countries" [22]. However, Bank of America has estimated that in the next 20 years there will be more than $20 trillion of material and financial growth in total investment in the global market, equal to about half of the current total market capitalization of the S&P 500, meaning relatively, current climate funding is still far from sufficient [23].

Various barriers can hinder investment. Examples include the immaturity of climate change policy frameworks and absence of investment policies that are stable, constraints on decision making within investor companies' fiduciary duties, perceptions of investors that returns on renewable infrastructure investments are too low and initial capital investment requirements too high, risk associated with uncertain and unproven technologies, high transaction costs or fees transaction costs, and lack of proven knowledge/technical advisement [24]. In addition, the most outstanding reason is the disinterest from politicians, who may fear losing their positions of power from carbon-relying groups such as automobile companies and people who tend to favor lower gasoline prices and more affordable vehicles. Fortunately, several policies and government interventions are being developed to reduce or manage barriers to investment [25]. These include the use of regulatory measures as well as public finance mechanisms (PFMs) and public-private partnerships (PPPs). The mandates and targets set for renewables, such as the European Union's 20% of final energy from renewable sources by 2020 goal, have also shown a somewhat positive result. Several other approaches have been utilized including subsidies or stricter government regulations. Though such governmental policies do aid in investing efforts, most of the funding must come from the private sector and public taxes [26].

*Government Subsidies & Programs:*

Subsidies are the most common and important policy to stimulate the development of the renewable energy industry. In the United States, the federal government has paid $145 billion for energy subsidies to support R&D for nuclear power ($85 billion) and fossil fuels ($60 billion) from 1950 to 2016. Comparatively, renewable energy technologies received a total of $34 billion [27]. In addition, most states have some financial incentives available to support or subsidize the installation of renewable energy equipment. Numerous types of subsidies have been implemented in European countries as well. The main policies being the Feed-in Tariff (FiT), the Feed-in Premium, and the Green Certificate (GC). To be successful, these policies usually include three key provisions: (1) guaranteed access to the grid; (2) stable, long-term purchase agreements (typically, about 15-20 years); and (3) payment levels based on the costs of renewable energy generation. It provides a fixed amount of money to be paid for renewable electricity production and an additional premium on top of the electricity market price [28]. The implementation of such policies can also be seen in other European countries: In Italy, the Gestore dei Servizi Energetici published some reports on renewable energy support policies. The Spanish authority, Comisión Nacional de Energía (CNE), produces information on energy policies and the British Office of Gas and Electricity Markets (OFGEM) publishes an annual report [29].

Though effective, government subsidies come with one negative aspect - its high cost. Government subsidies can be of significant financial burden as the increase in degree of attention to the environment can not only increase the price of energy products, it can also bring about the decline in output level of renewable energy enterprise, following from the premise that the size of the market remains unchanged in the short term [30]. From the perspective of promoting the development of renewable energy, when the environmental pollution caused by energy enterprises is slight, mixed forms of subsidy policy appear to provide the optimal path forward. Different subsidy policies have their own advantages and there are also relevant limitations in some respects. Hence, policymakers should make and implement reasonable policies to fit their own needs according to their own carefully considered situations and targets [31].

*Direct/Indirect Aid to Other Countries:*

Aiding developing world countries is a critical economic policy, but whose impacts will still be minimal if only a small portion of the globe is taking or can take the initiative. Thus, by providing aid in the form of investments, political and economic actions or otherwise, we can ensure these countries take the necessary steps to also combat climate change. As recently reported, and amid the COVID pandemic in 2021, the United States decided to rejoin international efforts against climate change. President Joe Biden has publicly stated he wants to re-establish U.S. leadership on climate. Doing so will require the United States to make an ambitious but achievable pledge and to assist other nations in doing the same. Nathan Hultman, a nonresident senior fellow in the Global Economy and Development program at Brookings, suggests that these subnational actors can share their skills and ambition with their counterparts abroad. Hultman also sees for the United States an opportunity to lead through its outsized role in the global financial sector. It can encourage greener investing by requiring disclosure of climate risks and support global efforts to finance emissions reduction and climate adaptation in developing countries [32].

Aid and interactions do not have to come directly from a given nation's political leadership. In fact, subnational actors with significant climate commitments represent roughly 70% of the U.S.'s GDP, which is roughly equivalent to the economy of China [33]. Using policy authorities at their disposal, many of which are significant, these actors have advanced climate action across multiple sectors and types of emission, including electricity, clean transportation, land use, methane, Chlorofluorocarbons, and more. Even outside of federal regulation and legislation, such policies are already driving significant reductions in U.S. emissions and could do more if expanded in line with recent trends [34]. As another example, over 600 local governments in the United States have developed climate action plans. While many of these municipalities are lagging in their efforts to meet their targets, some large cities (Los Angeles, New York City, and Durham, North Carolina, for example) have achieved significant reductions and have highly qualified organizations to demonstrate how such reductions can be achieved [35]. The United States can leverage its non-federal entities in its diplomatic efforts to support and bolster climate action around the world. Several examples of these economic solutions are listed in Table 3, below, along with their leading benefits and consequences of implementation, which are clearly stated for better comparison. Thus, U.S. cities, states, and businesses can collaborate with their counterparts in other countries to discuss opportunities and strategies, supported by U.S. diplomatic effort, as analyzed and recommended by Anthony F. Pipa, a senior fellow and Max Bouchet, a project manager and senior policy analyst, both of the Center for Sustainable Development, housed in the Global Economy and Development program at Brookings, in their brief for this series. Lastly, international perception of the U.S. domestic commitment is also important; the commitment must be seen as sufficiently ambitious to unlock other diplomatic opportunities available to the United States [33].

Table 3: A Summary of Leading Benefits and Consequences of Economic Solutions

|  | Leading Benefit | Leading Consequence |
|---|---|---|
| **Carbon Tax** | Government Revenue from Tax Returns | Lack of Consensus on Tax Rate |
| **Cap & Trade** | Market Flexibility & Satisfaction | Market Setbacks from Complementary Solutions |
| **Investments** | Secure Outcomes and Returns | Difficulty Motivating Investors |
| **Government Action** | Climate Industry Development | High Cost and Lowered Output |
| **International Aid** | Balanced Global Advancement | Difficulties Cooperating Internationally |

**Agricultural/Agroforestry**

*Afforestation/Reforestation:*

The Green Belt Movement (GBM), founded in 1977 by Wangari Maathai, planted 51 million trees in Kenya, and restored 850 hectares of the countryside in 2018. GBM is one of the many practices of afforestation and reforestation, a mitigation strategy through land use management, serving as reversal processes to forest and soil degradation, reducing the negative impacts on the hydrological systems, and aiming to bury $CO_2$ in the soil through photosynthesis. Afforestation is the introduction of trees and plants to clearings, wastelands, and arid, barren areas, while reforestation is the restoration of forests experiencing a significant decrease in tree population due to deforestation, wildfire, and other natural/human-made disasters.

As indicated in Table 4, below, afforestation and deforestation have comparably low costs ($0/tCO_2$ - $50/tCO_2$) with high carbon removal rates ($3.6 GtCO_2$ by 2050 & $7 GtCO_2$ by 2100) and capacities ($80 GtCO_2$ - $260 GtCO_2$ removed in total) that enable them to cover larger landmasses (70Mha – 500Mha) and extract abundant amounts of $CO_2$ with a relatively small budget. To put in perspective, an acre of matured trees absorbs 9.2 metric tons of $CO_2$ per year, yet the cost of planting one acre of trees does not exceed $1000 [36]. By introducing additional new trees and plants into this area, afforestation and reforestation help to prevent topsoil runoff and erosion as increasing the quantity of trees in near-barren lands pins down the soil with their interconnecting network of roots. Through transpiration,

torrential rainfall, sturdy underground watersheds and water tables are realized. An improved, cleaner environment assists in preserving endangered organisms and increasing the biodiversity of that area by providing a more supportive natural habitat. With new plants and trees introduced, the replenishment of fresher air dilutes the concentration of different respiratory diseases. In addition, an environment with less air pollution helps to shield society from illness and discomfort. Other benefits such as social cohesion, leisure activities, and the raising of awareness and education for future generations can all be observed through the implementation of afforestation and reforestation [37].

On the other hand, forest management, especially the creation of new forests on existing lands, can lead to the loss of land for urban development, habitats, biodiversity, agriculture, housing and other public infrastructure. Eco-tourism is also an unintended consequence of afforestation: those implemented solely for economic benefits and entertainment can bring more litter and destruction into forests and habitats rather than preserving them. Additionally, apart from natural disasters, expanding forest landmass increases the land value and scarcity which then contributes to an escalation of property prices [37].

Table 4: Afforestation/Reforestation and Forest Managements Data Analysis

| Work Cited: [38] [39] | Approximate time span (#/yr[a]) | Global annual $CO_2$ removal potential (GtCO$_2$[b]) | Global total $CO_2$ removal capacity (GtCO$_2$) | Global land mass required (Mha[c]) | Total cost for implementation of practice ($/tCO$_2$[d]) |
|---|---|---|---|---|---|
| *Afforestation/ Reforestation + Forest managements* | 1.9 billion | ≈ 3.6 (2050) ≈ 7 (2100) | [80, 260][e] | [70, 90] ∪[f] [350, 500] | [0, 50] |

a: Number of trees annually
b: Gigatons of $CO_2$ mitigated
c: Million hectare
d: Cost in United States Dollars (USD) for every ton of $CO_2$
e: "[x,y]" represents the domain and limitation of the variable from x to y
f: "[w,x] ∪ [y,z]" represents the conjunction of two (or more) domains, where it stands for "the limit is from w to x, and the limitation is also from y to z"

Afforestation and reforestation are widely executed by many states and regions. For example, the Republic of Korea (South Korea) has been conducting national reforestation program since 1961 [40], and the Korea Forest Service (KFS) has been intensively planting trees since the 1970s and 1980s. By 2008, 2960 million trees were planted across 1,080 thousand hectares of South and North Korean territory, bringing an alliance between these two bitter rivals on this matter. This strategy is ready to be put into large-scale carbon removal practice immediately with public approvals, helping to bring about a greener world for future generations.

*Bioenergy Carbon Capture and Storage (BECCS):*

In combating climate change, 23 bioenergy carbon capture and storage (BECCS) projects have been executed globally, with the majority in Europe and North America. Currently, 6 of the 23 remained in operation, "capturing $CO_2$ from ethanol bio-refinery plants and MSW (Municipal Solid Waste) recycling centers," and in 2019, 5 facilities are actively capturing ≈1.5 million tCO$_2$/y worldwide through BECCS technologies [41].

BECCS is one among an array of negative emissions technologies (NETs), a mitigation strategy. This technology entails the burning of biomasses such as forest woods and fast-growing crops (e.g., barley, wheat, corn, sugarcane, rice, and willow trees), from which bioenergy is generated (i.e., heat/electricity), and the $CO_2$ emitted from the process is captured into long-term underground storage [42].

Naturally, photosynthesis creates a carbon-neutral process. However, by seizing the escaping $CO_2$ before it reaches the atmosphere the net $CO_2$ emission can then be negative, decreasing $CO_2$ concentration in the atmosphere as represented in Table 5, BECCS can mitigate up to $1191 GtCO_2$ globally over the span of the technology's lifetime. Owing to its cultivation and burning of biomass, however, BECCS is largely limited by its high economic demand ($100 - $200 per $tCO_2$) and biomass availability (a demand of nearly 50% of Earth's agricultural landmass). Further, this option will decrease land-use availability for housing and crops, increase utilization of water, fertilizers, damage to local habitats, release of $CO_2$ from the soil, and create pipeline-related concerns due to $CO_2$ injection into geological reservoirs. These downsides can lead to an increase in food insecurities, displacements, biodiversity loss, shortage of water, soil carbon loss and leakage, seismic activity, and air/water pollution [39].

Table 5: Bioenergy Carbon Capture and Storage (BECCS) Data Analysis

| Work Cited: [38] [39] [43] | Approximate biomass/ bioenergy productivity (t/ha[a]) | Global annual $CO_2$ removal potential ($GtCO_2$[b]) | Global total $CO_2$ removal capacity ($GtCO_2$) | Global land mass required (Mha/$GtCO_2$) | Total cost for implementation of practice ($/$tCO_2$[c]) |
|---|---|---|---|---|---|
| **Bioenergy Carbon Capture and Storage (BECCS)** | [1.8, 25.1][d] | [0.5, 5] (2050)[e] [5, 10] (2100) | [0, 1191] | [31.7, 58.3] | [100, 200] |

a: Tons per hectare
b: Gigaton of $CO_2$
c: Cost in United State Dollars (USD) per tons of $CO_2$
d: "[x,y]" represents the domain and limitation of the variable from x to y
e: "(z)" represents the year said goal should be achieved

Like afforestation and reforestation, BECCS implementation is also distributed worldwide. However, governments should take into consideration that research, development, and demonstration (RD&D), along with life cycle analysis, agricultural policies, finance mechanisms, and cross-cutting considerations should be promoted and implemented for maximized benefits to be received from this program.

*Bioengineering (BE) of Crops/Genetically Modified Organisms (GMOs):*

With limited water and the degradation of soil health caused by climate change; food insecurity will continue to increase if no actions are taken. One of the possible adaptation strategies to climate change is the bioengineering of crops, where the alteration of crop's DNA, genes, and alleles allows farmers to yield crop productivity with smaller land areas. Different methods of bioengineering crops in agricultural practices are available according to current technology: traditional breeding, mutagenesis, RNA interference, transgenesis, and gene editing.

*Traditional breeding*, scientifically established by Gregor Mendel in the 1860s, focuses on the selection of desirable alleles and cross breeding these selected crops together to produce offspring that combines both beneficial traits while minimizing disadvantages against its environment. This method, dating back approximately 9,000 to 11,000 years, does not require further research and testing for large-scale implementation/organic use, and can affect up to 10,000-300,000 genes in total. *Mutagenesis*, invented in 1983 by Kary B. Mullis, is a technique using chemicals and radiation that efficiently detects and escalates a targeted genome/DNA sequence, amplifying the desired genes without cloning. Although mutagenesis does not require testing for implementation and is approved for organic uses, it remains extremely unpredictable, therefore creating uncertainty on the number of genes it can affect during its process. *RNA interference*, discovered by Andrew Fire and Craig Mello in 1998, presents itself as a mechanism that can inhibit certain gene expressions by degrading mRNA from that specifically chosen gene and neutralizing the mRNA. Yet it can only be conducted under the condition that the RNA molecules appear as double-stranded pairs. With future testing required for application and the ability to affect 1 to 2 genes, this method is not well-established for organic use, though RNA interference may be able to effectively reduce certain traits of crop in the future. *Transgenesis*, developed in 1973 by Herbert Boyer and Stanley Cohen, involves the practice of transferring a section of the desired gene(s) from one organism to another in a specific location to promote chosen traits. This method requires further testing for implementation outside of organic use, it can affect approximately 1 to 3 genes during each transferring process. With more technological development, transgenesis would allow crops that had been genetically modified to pass on their altered traits to future generations. Ultimately, there are *gene-editing* methods such as the Clustered Regularly Interspaced Short Palindromic Repeats (CRISPR) technique, established by Emmanuelle Charpentier and Jennifer Doudna in the early 2010s, that through identification of the "problematic" genes conduct "operations" to alter those genes to the desired form and script. Even with testing required and an unknown certainty of organic utilization, it can accurately affect at least 1 to 3 genes when applied, ending with a result that is more specific, controlled, and predictable [44].

With the bioengineering of crops, the products can become much more resilient to the degrading environment on Earth caused by climate change, maintaining increases in overall output from strengthened protection against extreme weather conditions (e.g., drought, flood, storm, strong wind, etc.). As a result, a greater portion of these crops will be saved from both unpredictable natural and human-caused disasters, expanding the suitable soil range for the crops to be planted (i.e., ~10% increase of global arable land), and reducing the use of chemicals/fertilizers (e.g., pesticides) and tillage (i.e., the process of turning the soil in the field), which in turn improves soil preparation efficiency and reduces greenhouse gas emissions. With better security, larger agricultural land dedication, and higher production, food security can increase dramatically, supporting the growing global population and leaving fewer people to contend with insufficient daily nutrition. However, not only will developing the altered seeds require large monetary investments, but unpredictability in seed qualities raises the investment rate further, along with the complicating factor of often being mixed with regular seeds that do not match pricewise [45].

This adaptation technology has already been implemented by some countries and regions. Unfortunately, the bioengineering of crops still requires much more research to decrease the unpredictability of the genes that are being modified, and to increase the effectiveness of the modified genes to maximize the advantages brought by the GMOs. In addition, better education and regulations for the farmers that utilize these GMOs should be established to raise awareness of the negative effects of all aspects brought by climate change and address the concerns of GMOs [45].

*Irrigation Systems:*

Worldwide, average agricultural water utilization practices equate to more than 70% of global freshwater consumption annually [46], while approximately 40% of that water consumption is wasted through primitive irrigation methods and failure of resource management (i.e., "poor irrigation systems[/transportation], evaporation, [water runoff,] and overall poor water management") [47]. Correspondingly, some fruits and vegetables require more water usage than others. To put in perspective, to grow one pound of wheat uses 130 gallons of water while the same quantity of coffee requires 2,500 gallons of water. In many cases, the 40% of water not actually used for its intended task was returned to the environment rather than being used elsewhere, resulting in more input of money, time, and energy consumption to re-acquire and redistribute this water [47]. Therefore, to better manage water distribution overall, better irrigation systems are needed for large-scale implementation.

Like the bioengineering of crops, there are many methods for this adaptation strategy from surface irrigation (i.e., traditional water delivery systems) and sprinkler irrigation to drip and airdrop irrigation systems. The drip irrigation system (i.e., micro/low-flow/low-volume/trickle irrigation system), first introduced by Simcha Blass and Kibbutz Hatzerim in 1959, creates a "dripping" system that maintains soil moisture at a fixed level through water-emitting technologies applying droplets and small streams of water to the soil surface/plant roots Water consumption is tightly controlled at up to 90% water-use efficiency while providing a much more effective and efficient way of applying chemicals and fertilizers to the soil. One example of the drip irrigation system is subsurface irrigation (SDI), which similarly irrigates the crop as the drip irrigation system from underground and within the plant root zone for better water delivery accuracy and overall management. With the more developed technologies - e.g., "pumps/pressurized water system, filtration systems, nutrients application system, backwash controllers, pressure control valves (i.e., pressure regulators), pipes (including main pipelines and branching tubes), control/safety valves, poly fittings, accessories, and emitters" [45], the accuracy of water usage can be greatly improved. By reducing deep percolation/evaporation water run-off to near zero decreases in production input, diseases, and the unpredictability of crop growth result while increasing the yield and quality of the finished crops. Furthermore, the drip irrigation systems can be automated and applied across many climates, conditions, and soils (e.g., salinity, sandy, drought, terrains) that other irrigation systems may not adapt to, supporting a wider variety of permanent/non-permanent crops, fruits, and vegetables. The biggest concern regarding the drip irrigation system is the cost. Due to the many instruments needed for this practice, the initial cost of implementation ($800 to $2,500/hectare) can be considerably high. However, in maintaining the practice, fluctuations of the cost may be affected by unpredictable rainfall, climate/soil conditions, damage to wildlife, and the shifting of piping/instrumentation positions.

A more recently developed irrigation system, invented in 2011 by Edward Linacre, is the airdrop irrigation system. This technology essentially harvests $H_2O$ molecules or moisture droplets from the air through a turbine that drives and cools the air to that of the underground space in a condensation process until it reaches 100% humidity, resulting in condensate formation. The produced water, stored in an underground tank, is then pumped to the roots of the plants during the watering process. Because "the airdrop irrigation system is a low-tech, self-sufficient solar-powered solution," [48] it is suitable for arid and semi-arid land where water shortage presents as a recurring problem, so less water can be used in a more cost-effective manner.

Most irrigation system types are currently widely implemented. However, with better economic management and public awareness, improved technologies and instruments can be applied to integrate the overall benefits provided by these systems. By implementing systems such as the drip and airdrop irrigation systems, water usage can be substantially decreased, while the creation of artificial ponds, lakes, and reservoirs can supply farmers with a constant water supply, relieving otherwise persistent water shortage pressures in some regions.

## **Atmospheric/Astronomical**

*Carbon Capture, Utilization, and Storage:*

While technologies to decrease greenhouse gas emissions are vital to meeting climate goals, negative emission technologies must also be analyzed and considered to formulate the most optimal combination of strategies. Carbon capture, utilization, and storage (CCUS) is a type of negative emission technology (NET) designed to chemically capture $CO_2$ from the atmosphere, concentrate it, and inject it underground or into a storage reservoir.

CCUS systems capture $CO_2$ from either the source of emission or from the atmosphere via direct air capture (DAC) and permanently stores the greenhouse gas underground. Globally, approximately 8 gigatonnes of $CO_2$ must be removed annually to stay within the goals mentioned previously corresponding to a relatively safe range of increasing temperature. The low-end cost of $100 per metric ton of $CO_2$ captured and stored is higher than most other mitigation technologies, mainly due to the high levels of energy needed to separate $CO_2$ from the solutes or sorbents used in the capture of the GHG during the chemical process. In addition, captured $CO_2$ as a commodity does not attract a large market. However, there have been recent technological developments such as enhanced oil recovery and synthetic aggregates that could provide a large enough market to lower the cost barrier of CCUS. This negative emission technology requires very little land overall and does not require such land to be arable - one of its major advantages compared to other mitigation technologies. The water usage associated with CCUS depends on the humidity and ambient temperature of the environment [49]. Designating CCUS plants in cooler, more humid climates can minimize the amount of water lost due to evaporation, thus reducing the amount of water needed in the process.

As greenhouse gas emissions rapidly increase, it becomes clear that simply reducing emissions will not be enough to reduce the effects of global warming; instead, climate change will only be fully moderated by removing $CO_2$ directly from the atmosphere in combination with converting to renewable energy. In fact, the IPCC states that "all pathways that limit global warming to 1.5°C with limited or no overshoot project the use of carbon dioxide removal" [50], emphasizing the importance of implementing carbon capture, utilization, and storage. One of the major benefits of CCUS is its practical land requirements for the system, which would lessen negative impacts on local food production or other land uses. Compared with other mitigation technologies, CCUS plants require much less space. Captured $CO_2$ can also be sold or recycled to bring in revenue and help lessen the cost of carbon capture such as being integrated into synthetic fuels or building insulation.

However, as shown in Table 6, carbon capture, utilization, and storage systems require substantial amounts of energy to power equipment and regulate the rate of carbon capture (i.e., 2000kWh/tonne of $CO_2$ removed). One study found that the energy needed to provide enough power and heat to the process to meet the Paris Agreement objectives was approximately a quarter of global energy supplies by 2100 [51]. Moreover, as indicated in Table 6, the process of CCUS is very expensive, in the range of $1 trillion - $10 trillion/year for removal of 10 gigatonnes $CO_2$ per year. As previously stated, the cost of adhering to the 1.5°C pathway would run into the trillions of dollars. Accordingly, while CCUS is crucial to implement it should not be heavily relied upon. Depending on CCUS to combat climate change can also lead to the misguided belief that emissions from burning fossil fuels can be offset with GHG removal technologies, when CCUS is expensive and itself consumes an enormous amount of energy. This circumstance creates a challenging GHG balance and would demand zero, or at least low, GHG emitting energy sources to be employed in powering CCUS facilities.

As of 2021, Climeworks – a notable company specializing in carbon air capture technology – operates three CCUS plants that capture approximately 1,100 tonnes of $CO_2$ per year. Climeworks plants are 90% efficient and emit ≈10 kg for every 100 kg of $CO_2$ removed from the atmosphere. After capturing $CO_2$, Climeworks either sequesters the collected greenhouse gas underground or sells it for commercial purposes. Although Climeworks has yet to reach profitability, there is still much reason to believe that they, along with many other similar for-profit businesses or even government-owned plants, can pull vast amounts of $CO_2$ out of the atmosphere and bury it underground while selling enough $CO_2$ to provide an offset to their operational costs to continue. The most optimistic scenario is one in which a virtuous circle occurs: when copious amounts of $CO_2$ is produced and attracts a larger market that can leverage the economics of scale, thus driving the cycle.

Table 6: Carbon Capture, Utilization, and Storage Data Analysis

| Global land mass required (km$^2$) | Cost for implementation of practice ($/tonne removed) | Total cost for implementation to remove 10 gigatonnes $CO_2$ per year (see Introduction) | Energy consumption (kWh[a]/tonne removed) | Water usage (tonne/tonne removed) |
|---|---|---|---|---|
| 400 km$^2$ to 24,700 km$^2$ per Gt (non-arable) | $100 - $1,000/tonne removed | $1 trillion - $10 trillion/year | 2,000 kWh/tonne of $CO_2$ removed | 1 - 7 tonnes/tonne of $CO_2$ removed |

a: Kilowatt hour (3,600 kilojoules)

*Stratospheric Aerosol Injection:*

In June 1991, a volcano located in the Philippines (Mount Pinatubo) erupted, explosively ejecting about 20 million tons of sulfur dioxide into the atmosphere which, after forming particulates, reflected substantial amounts of sunlight back into space that would have otherwise reached the Earth's surface. As the contents of Pinatubo volcanic plumes became distributed across the planet, major cooling effects resulted. This eruption lowered the global temperature by nearly 0.6 °C. Stratospheric aerosol injection (SAI) aims to mimic this cooling effect by spraying large quantities of reflective particles into the stratosphere. Because aerosol particles scatter and absorb sunlight, they can greatly influence plans to cool the climate. However, due to its many shortcomings, this proposed type of climate engineering is currently only theoretical.

Chiefly, the obvious benefit of SAI methodology involves the rapid cooling of Earth. The amount of cooling and the duration of its effects depend on the type, amount, and persistence of the aerosols to remain suspended. Possible particle types range from sulfur dioxide (commonly used as sprayed reflective particles) to finely powdered salt or calcium. Unlike marine sky brightening (discussed later in this section), SAI is not deployed in the atmosphere but rather in the stratosphere, which does not contain heavy rain clouds that could quickly disperse these otherwise pollutants. Thus, it is likely that heavy influxes of aerosol particles could remain in the stratosphere for a longer amount of time until removal by natural chemical processes and atmospheric. If efforts were to be implemented successfully, most climate change mitigation objectives would follow, including the reduction or reversal of land/sea ice sheet melting, an increase in plant productivity, reduction/reversal of sea-level rise, and an increase in terrestrial $CO_2$ sink from enhanced sequestration in soil and oceans.

On the contrary, SAI is dismissed by many members of the scientific community because of many potential negative consequences upon implementation. While SAI would lower global temperatures, droughts in certain continents would still likely ensue, if not worsen. According to the Geoengineering Model Intercomparison Project, temperatures in tropical areas would cool, yet areas with higher latitudes would warm, as well as cause increased extreme climates and ice sheet melting. Additionally, climate change-related problems such as ocean acidification from $CO_2$ forming carbonic acid would not be addressed and resolved by SAI since this technique can only mitigate surface climate issues. A variety of atmospheric impediments would also arise from SAI, for example, solar power and ground-based optical astronomy would be greatly hindered using this mitigation strategy. Commercial/military control of this technology should also be mentioned as possible consequences of SAI. In addition, international conflicts would be extremely difficult to avoid because the implementation of SAI would require the agreement of each country, and withdrawal from the agreement at any time could cause the entire operation to fail. This could lead to the termination effect, which would have disastrous consequences once this grand-scale geoengineering strategy is paused. "If geoengineering were halted all at once, there would be rapid temperature and precipitation increases at 5–10 times the rates from gradual global warming" [52]. Lastly, just as other mitigation strategies such as space-based mirrors and direct air capture can introduce a moral hazard, success with SAI may cause global populations and governments to increase support towards these temporary technologies and reduce funds for permanent solutions such as renewable energy and better agricultural practices. Thus, it is vital for government leaders and policymakers to acknowledge that these solutions are short-term, flawed, and only considered as temporary relief from the repercussions of global warming.

After examination of its many disadvantages in practice, SAI is suggested to be best regarded as only a theory in need of significantly more research rather than having real application potential in the near term. Other solutions that are more long-term and do not involve nearly so many negative consequences should be evaluated further and considered instead. If SAI is implemented, policymakers must remember that SAI is not and cannot be a substitute for permanent mitigation strategies.

*Marine Sky Brightening:*

The basic idea of marine sky/cloud brightening (MSB) is to enhance cloud reflectivity by cloud seeding with seawater droplets or with other synthesized chemicals. Seawater is sprayed into the air to inject salt into the clouds, increasing albedo (fraction of incident sunlight reflection back into space), thereby aiming to offset climate change. This type of solar radiation management (SRM) would require enough salt crystals to ensure an effective reflection rate while also being small enough in size to not promote precipitation. If implemented successfully, cooling effects would promptly follow and could be effective in mitigating global warming.

The impacts of marine sky brightening would be immediate and reversible in the short term. Compared to other SRM techniques, marine sky brightening is considerably more financially feasible. According to a report published in 2008, in addition to approximately £50 million for research, development, and tooling, "50 spray vessels costing approximately £1–2 million [$2 million] each could cancel the thermal effects of a 1-year increase in world $CO_2$", adding up to nearly £50-110 million to offset the thermal damage done by 1 year of $CO_2$ increase [53]. By comparison, space-based mirrors require between one to ten trillion dollars for effectiveness. Furthermore, MSB also allows for the localization of solar radiation protection. This technology can be directed to shield specific regions such as areas with ice sheets that are at greater risk due to global warming.

Like deploying sunshade configurations in the atmosphere, MSB also concerns international law and politics. In particular, the United Nations Convention on the Law of the Sea (UNCLOS) states that parties are obligated to "protect and preserve the marine environment" from any polluting source. Whether MSB particles would be considered a polluting source remains unclear, and any negative effects of MSB could lead to immediate violation of the law. Moreover, a limited understanding of the complex nature of clouds could lead to unexpected consequences. Large-scale climate patterns and precipitation could be greatly affected, though climate experts suggest that SAI technology and usual climate change patterns would result in more drastic weather changes.

Enhancing cloud reflectivity has the most impact (both beneficial and detrimental) on local and regional precipitation, temperatures, and run-off. Thus, compared to SAI technology, MSB can localize its reflective effects and is roughly less expensive. While both should be considered as temporary fixes, MSB may be more moderate for known adverse effects while SAI involves several unknown effects. The brightness of clouds and reflectivity rates drop dramatically after a few days of cessation of the technology; thereby making MSB easily controllable and an ideal temporary relief solution. It is important to note, however, that a sharp halt of activity could lead to the termination effect, and the reduced carbon sink could lead to significantly warmer temperatures, like the usage case of SAI. Marine cloud brightening, although able to produce a reductive effect on both regional and global warming, will likely cause its own changes to climate, and constant cloud assessments and modifications will be required to ensure there are no serious adverse effects of MSB.

*Space-Based Mirrors:*

In addition to efforts for reducing GHG emissions and storing away GHG, alternative methods to counteract global warming have been theorized and researched for potential deployment. Among these include a newly developed yet promising field: space-based mirrors. Like the concepts of stratospheric aerosol injection and marine sky brightening, this idea aims to reflect solar energy away from Earth. It is important to note that this strategy, along with other solar radiation management techniques, should be viewed as a last resort rather than a continuously implemented policy. This is due, in part, to some of the disadvantages regarding these rapid GHG reduction processes that will be discussed later.

Although the deployment of large-scale space-based mirrors may pose too monumental a task to be practical, sunshade configurations have been viewed to be one of the most efficient methods in solving climate change [54]. Deploying large orbital sunshades allows for the concentration of specific areas that are most directly impacted by the effects of climate change. By erecting shields to prevent overheating in those areas, local and regional environments that face the greatest dangers can be temporarily rescued until other, more permanent solutions are implemented.

However, only by reducing GHG emissions and addressing the excess GHG already existing in our atmosphere and oceans can a permanently stable state of life on Earth be achieved. Future generations cannot rely on simply resisting climate change without addressing its root causes, and future implementation of space shades followed by their success might lead the public to demand more of the same, eventually resulting in the dependence of these short-term, "back end of the pipe" relief measures. Ocean acidification, among other environmental issues caused by excessive GHGs in the atmosphere, would remain entirely unsolved by the blockage of sunlight, whether via mirrors or reflective particles. In addition, the present economic feasibility of this solution is low unless stronger motivations and funding for SAI emerges. Some estimations place the cost of space transportation and construction to be between one and ten trillion dollars. Lastly, adverse effects such as unintended influences on Earth's various natural cycles and cultivation of crops are also shortcomings to carefully consider.

While computer simulations have demonstrated that space-based mirrors are theoretically successful, experiments have not yet been conducted at a scale large enough to ensure the safety and effectiveness of this mitigation strategy in the real world. In addition, global consensus must be achieved before implementation, otherwise negative results occurring in some countries while success in others may prompt political blame and, in the worst case, global warfare. As per the general view of the scientific community, it would be most optimal to continue regarding solar radiation management techniques (SRM) as a last resort and undertake extreme caution during any implementation efforts.

**Geological**

*Geologic Reservoir Sequestration:*

Geologic reservoir sequestration, or geological sequestration, is a mitigation strategy used after $CO_2$ is captured "at the point of emission" from industrial methods. This is done by storing captured $CO_2$ "in deep underground geological formations" through physical or chemical implementations. Like the storing process in the Bioenergy Carbon Capture and Storage (BECCS) solution, the $CO_2$ is physically stored "within a cavity in the rock underground," regardless of whether these geological structures are "large man-made cavities" or "the pore space present within rock formations" [55]. Differentiated from BECCS, $CO_2$ sequestered into geologic reservoirs is usually "pressurized until it becomes a liquid, and then…injected into porous rock formations in geologic basins", and this process of carbon storage, also known as tertiary recovery, plays an important part in enhanced oil recovery [53]. Other methods of storing fully oxidized carbon involve the transformation of $CO_2$, such as "dissolving $CO_2$ in underground water or reservoir oil", "adsorption trapping", "decomposing $CO_2$ into its ionic components", and chemically combining and attaching these captured carbons with other underground substances by "locking $CO_2$ into a stable mineral precipitate" [55]. Large volumes of these types of formations can be found in the U.S.'s coastal plains regions (e.g., "The coastal basin from Texas to Georgia. ...accounts for 2,000 metric gigatons, or 65[%], of the storage potential" [56], and the abundance of carbon storage capacity in geologic reservoirs can also be observed in Table 7, where the global capacity of carbon storage ranges from 5,000$GtCO_2$ to 25,000$GtCO_2$. Therefore, it is crucial for organizations to pinpoint the most optimal locations for such implementation, such as "mature oil and natural gas reservoirs [,] oil and gas-rich organic shale [,] uneconomic coalbeds [,] deep aquifers saturated with brackish water or brine (saline) [,] salt caverns [, and] basalt formations," [55] can help to ensure the process proceeds smoothly.

One of the most obvious benefits of geologic reservoir sequestration is the improvement in atmospheric concentrations of $CO_2$. By trapping the $CO_2$ before it reaches the atmosphere, $CO_2$ loading will be reduced therefore slowing down the growth rate of greenhouse gasses. This process, however, may generate a larger consumption of fossil fuels if no cleaner energy sources are broadly adopted, which leads to one of the main concerns regarding geologic reservoir sequestration - the location and transportation of $CO_2$ as it affects the overall balance of energy and $CO_2$. Due to the locations of most "oil sands and coal-burning electrical plants" being situated away from the suitable geological areas for carbon injection, $CO_2$ must be transported through pipes or on trucks over long distances to be stored underground [55]. The transportation process entails extra costs and energy, and if not done carefully can emit a considerable amount of $CO_2$ itself which undermines the strategy's intent. To put in perspective, approximately $88.90 is required to transport 1 million tonnes per annum of $CO_2$ ($MtpaCO_2$) over 500 miles and "assumes extra monitoring requirements for $CO_2$ storage" [57]. In addition, an increase in energy and resource consumption can be observed through the construction and operation of such facilities, bringing further concerns both economically and environmentally to the surface. The number of carbon injections is ultimately limited to prevent increasing the probability of natural disasters such as earthquakes. Current EPA underground injection control programs, such as the Maximum Allowable Surface Injection Pressure (MASIP), establish regulations for carbon injections based on "calculated, testable, and well documented" pressure requirements that will prevent unintended formation fracturing which may arise during the process of injections. Other substantive risks include fracking which would potentially lead to brine water leakage and result in freshwater contamination. This, in turn, may also affect how the strategy is perceived by the government and the public [58].

As a contemporary of BECCS, geologic reservoir sequestration has been implemented in only a few instances and is still largely in its developmental stage. Some examples where geological sequestration is used include "offshore natural gas production" and to "boost production from oil fields by displacing trapped oil and gas". Similar uses of this strategy can be further adopted as the technology more fully develops (e.g., carbon transportation, pipe leakage prevention, injection methods/architectures, etc.), increasing its carbon capture potential and reducing the costs and landmass requirements, as shown in Table 7 [59]. Geological sequestration is largely interconnected with many other mitigation technologies, so the increased implementation of others is likely to lead to an expansion of this practice as well.

Table 7: Geologic Reservoir Sequestration Data Analysis

| **Work Cited:** [38] | Approximate $CO_2$ sequestered in depleted oil reservoirs | Global annual $CO_2$ removal potential ($GtCO_2$) | Global total $CO_2$ removal capacity ($GtCO_2$)[a] | Global land mass required (Mt) to store $CO_2$ ($km^2$) | Total cost for implementation of practice ($/tCO_2$) |
|---|---|---|---|---|---|
| *Geologic Reservoir Sequestration* | 30 $GtCO_2$ | ≈ 35 | 5,000 – 25,000 | 50 - 100 Mt ≈ 100 $km^2$ | 7 - 13 |

a: Gigaton of $CO_2$.

*Soil Carbon Sequestration:*

Throughout human history most anthropogenic soil alterations usually resulted in a degradation of up to 50% to 70% of soil carbon storage and decreased more than 840 $GtCO_2$ of soil carbon. For example, forests were converted into farms or croplands, and farms were replaced by industrial factories or cities, etc. Therefore, soil carbon sequestration serves as a reversal process of these and other carbon-depleting practices, restoring the soil using plants best suited to the land, transforming infertile soil back to its initial generative states and re-introduces "the chemicals that inhibit the mycorrhizal and microbial interactions that store carbon" [60]. "Launched by France on December 1st, 2015, at the COP 21," the 4 per 1000 initiative is one of the many soil carbon sequestration organizations and initiatives [61]. Intending to increase plant and soil (top 30 – 40 cm) carbon absorption and storage by 4% every year through afforestation and other agroecological practices, the initiative not only hopes to improve soil carbon storage but also food security and agricultural adaptation under climate change. Other practices such as changes in agricultural methods and restoration of forests, grasslands, and wetlands can all yield increases in soil carbon storage.

The main benefits brought through soil carbon sequestration are that a healthier soil obtains a stronger defense against challenges brought by climate change such as drought, flood, and heavy rainfall, and by requiring fewer fertilizers to be used, is economically, ecologically, and environmentally less of a burden, where indicated in Table 8, it demands a minimal amount of cost (cost of technique does not exceed $100/tCO_2$ while it is able to cost as low as $0/tCO_2$) and soil while capable of storing a significant amount of carbon dioxide in the soil (soil carbon storage is able to increase to as high as 130$GtCO_2$ globally by 2100) In addition, by improving and restoring the health of the soil, afforestation and reforestation can encourage an increase in agricultural productivity.

However, soil carbon storage is limited by the soil's natural capacity at a given location, so residents and farmers are encouraged to better understand the details of new techniques and their role in increasing soil carbon storage. Transitions from one agricultural technique to another requires time and money, so providing a considerable amount of financial support to the people involved can efficiently improve the smoothness of this transition. In addition, the composition of soils varies worldwide, so to truly understand which species of plant or crop and farming techniques are best for a specific region and its type of soil requires a large amount of research. By encouraging research, different areas will have a better understanding of the particulars of their soil and accordingly, will be better equipped to maximize improvements by implementing the most optimal techniques for their soil rather than merely planting more trees.

Similarly, "blue carbon" (discussed further in the following section) serves the same purpose as soil carbon sequestration, only here sequestration is implemented in coastal and other regions involving bodies of water (e.g., mangroves, tidal marshlands, seagrass beds, and other tidal or saltwater wetlands).

Table 8: Soil Carbon Sequestration Data Analysis

| Work Cited: [50] [62] | Measurable amount of soil for $CO_2$ (cm) | Global annual $CO_2$ removal potential (GtCO$_2$)[a] | Global total $CO_2$ removal capacity (GtCO$_2$)[a] | Global soil carbon storage capability (tCO$_2$/acre) | Total cost for implementation of practice ($/tCO$_2$) |
|---|---|---|---|---|---|
| *Soil carbon sequestration* | $(15 - 40)$[b] | $(1 - 5)$[b]: 2050[c] | $(104 - 130)$[b]: 2100[c] | $\approx 8$ | $(0 - 100)$[b] |

a: Gigaton of $CO_2$
b: "x - y" represents the domain and limitation of the variable from x to y
c: "(z)" represents the year said goal should be achieved

**Coastal/Oceanic**

For coastal and oceanic areas, $CO_2$ removal techniques are separated into four general sub-sections: ecosystem restoration (e.g., mangrove/seaweed/wetland restoration, marine permaculture, and restocking of whale populations), ocean fertilization (e.g., iron/nitrogen/phosphorus fertilization and artificial upwelling/downwelling), modification of ocean chemistry (e.g., ocean alkalinity enhancement and seawater $CO_2$ stripping), and $CO_2$ storage (e.g., seabed/sub-seabed storage of $CO_2$ capture on land, and deep-sea storage of crop waste/macroalgae deposition). Although not all the listed solutions will be explored, some of the most optimal and beneficial sub-sections and solutions are included below (e.g., ocean alkalinity enhancement, ocean fertilization, and enhanced ocean productivity) [63].

*Ocean Alkalinity Enhancement:*

Affected by the worsening of climate change and global warming conditions, the ocean presents many concerns such as sea level and temperature rise, melting of the polar ice caps, and ecosystem and bio-habitat imbalances. This has caused the escalation of ocean acidity, disrupting the complex food web of the oceans. Utilizing the vast material for carbon capture and storage provided by the ocean, a mitigation strategy can be implemented that can chemically lock away $CO_2$ from the atmosphere in the ocean basin for hundreds and possibly thousands of years. Currently, ocean basins worldwide naturally hold roughly 39,000 GtCO2, while Earth's atmosphere 412 parts per million (ppm) – about 50% above pre-industrial level. Alkalinity can neutralize ocean acidity through alkaline minerals (e.g., limestone and basalt) weathering and eroding, from which the alkaline minerals extract hydrogen ions ($H^+$) from the ocean basin to drive up the basicity of the seawater. This restoration of equilibrium can be achieved, albeit over geologically long periods of time, via the reaction where "[the] chemical change shifts the carbonate chemistry equilibrium from dissolved $CO_2$ and carbonic acid ($H_2O + CO_2$) to bicarbonate ($HCO_3^-$) and carbonate ($CO_3^{2-}$) ions," so "[t]he conversion of dissolv[ing] $CO_2$ into bicarbonate [would be able to] create a $CO_2$-deficiency in surface waters, thereby pulling more atmospheric $CO_2$ into the ocean". In another $CO_2$ mitigation path, carbonic acid, produced when $CO_2$ reacts with the seawater, is broken into hydrogen ions and bicarbonate ions and calcifying organisms then transform these bicarbonate ions into calcium carbonate as the chief components of their shells and skeletons. As those organisms expire, they bury the calcium carbonate they carry within themselves on the ocean floor and the $CO_2$ is locked away in the form of minerals. Throughout recent human history, this natural process alone managed to "absorb approximately thirty percent of [the] anthropogenic carbon $CO_2$ emissions since the beginning of the Industrial Revolution" [60]. Regardless of the natural process of ocean acidity neutralization and carbon storage might exceed the human survival timeline itself, this constitutes only a minimal force in combating climate change. Therefore, artificial ocean alkalinity enhancement such as the "accelerated weathering of alkaline rock", "addition of manufactured alkalinity products", and molecular pumps are recommended for consideration when adapting this negative emission technology to increase this process's efficiency and effectiveness in carbon extraction and sequestration [64].

Ocean alkalinity enhancement can be put into practice by accelerating the natural process of locking $CO_2$ and ocean acidity in the basin in a variety of ways, requiring ≈30% of the ocean body. Table 9 provides a detailed description on the potential and requirements for ocean alkalinity enhancement technology, demonstrating its benefits and hurdles to implementation. Two ocean alkalinity enhancement methods will be explored in this section. The first strategy is to use controlled accelerated weathering reactors that combine crushed limestone, extracted seawater, and $CO_2$-rich flue gas to separate the acidic seawater and create an alkalized seawater solution that is then injected back into the ocean and locked away in the deep ocean's carbon vault. The second strategy is to insert finely ground alkaline rocks (e.g., limestone ⇒ lime, and silicate-rich rocks) into the ocean floor, thereby promoting and advancing the natural geological cycle of securing $CO_2$ in the ocean basin. Both strategies mimic oceanic carbon extraction processes which have been occurring over the past few billions of years, but are accelerated in this practice to greatly compress its naturally long timescale [64].

When ocean alkalinity enhancement is implemented to scale, this in proportion to the threats posed by climate change, a significant amount of $CO_2$ can be mitigated [65]: electrochemical weathering, a technology that pumps seawater through an electrochemical system, rearranging the water and salt molecules to produce two separate solutions: acidic and basic. The acidic solution is removed and can be sampled for scientific research enabling better ocean alkalinity enhancement methods, while the basic solution is injected back into the ocean to neutralize ocean acidity and increase the ocean's carbon extraction ability [64]. For this method specifically, valuable by-products such as hydrogen and oxygen gas, silica, and nickel/iron hydroxides are created as a source of energy. Other methods, such as the utilization of silicate-rich minerals (e.g., olivine), produce carbonate sediments which are discharged into the sea water where they release iron and silica, fertilizing the ocean's biodiversity in various scales. Moreover, as described in Table 9, this method alone can mitigate 12% of the global energy-related $CO_2$ emissions annually (roughly $2.5 GtCO_2 – 2.9 GtCO_2$), hence ocean alkalinity enhancement methods can capture and sequester vast amounts of $CO_2$ from the atmosphere for an extremely long duration lasting up to hundreds of thousands of years.

The biggest concern regarding ocean alkalinity enhancement technologies is the uncertainties of the processes. Biogeochemical side effects such as alteration of ocean chemistry and damage to the marine ecosystem are likely to be introduced by this mitigation strategy, where such changes can intensify the vulnerability of biodiversity, food security, resident health, water quality, etc., and possibly disrupt and degrade local and even regional economics [66]. The root cause of biogeochemical side effects is largely found in the heavy metals embedded in the alkaline materials that are dumped into the ocean, which can become widespread among the oceanic food chains, and the extensive mining of alkaline raw materials, raising environmental, societal, and local health concerns along with these processes having a comparably high level of energy consumption [67].

Table 9: Ocean Alkalinity Enhancement Data Analysis

| Work Cited: [63] [66] [68] [69] | Proportion of carbon (t) $\Longleftrightarrow$ material (t[a]) | Global annual $CO_2$ removal potential (GtCO$_2$)[b] | Global total anthropogenic activity $CO_2$ extraction (GtCO$_2$) | Global land mass required (% of Earth's surface) | Total cost for implementation of practice ($/tCO$_2$)[c] |
|---|---|---|---|---|---|
| *Ocean alkalinity enhancement* | $1 tCO_2$ = 1 - 3.5 | ≥ 2.5 - 2.9 | ≈ 67 | ≥ 70.9% | 5 - 160 |

a: Ton
b: Gigaton of $CO_2$
c: Cost in United States Dollars (USD) per tons of $CO_2$

Despite developments in research on ocean alkalinization and ocean chemistry associated techniques, ocean alkalinity enhancement technology still resides at an early theoretical level. Therefore, the government should regard research, development, operational regulations, environmental restrictions, and social sustainability be applied and carried out with the implementation of the ocean alkalinity enhancement to promote the beneficial factors of this technology while not neglecting the negative effects it brings [66].

*Ocean Fertilization (OF):*

Ocean fertilization (OF) utilizes the alteration of geoengineering on the ocean surface and ecosystem by adding nutrients (e.g., iron) to the upper layer of the ocean (i.e., euphotic zone) to increase the phytoplankton population and activity, increasing the ocean's carbon extraction efficiency and capacity. Because the marine carbon and nutrient cycle is considerably complex, to safely implement this mitigation strategy a detailed understanding of marine biology and the carbon/nutrient cycle must be obtained.

Vertical characterization of the ocean can be roughly described as four layers: surface ocean (i.e., euphotic zone: 0 - 100 m), twilight zone (i.e., mesopelagic zone: 100 - 1000 m), deep ocean (≥3700 m) and the seafloor (i.e., benthic zone). Activities and exchanges of carbon and nutrients thrive between the first three layers, while the seafloor contains mostly reactive sediments and the burial of $CO_2$. Starting from the first layer, the large phytoplankton resides at the ocean surface, consuming $CO_2$ and atmospheric depositions (e.g., iron [Fe] and nitrogen [N]) through photosynthesis, which are then consumed by the bacteria, viruses, and zooplankton in the microbial loop. Zooplankton, who also prey on the small phytoplankton and microzooplankton populations, are then captured by a higher trophic level of marine organisms, who are then devoured by the predators such as birds and fish. Through aquatic respiration, marine organisms extract the dissolved oxygen from the ocean water and excrete metabolic waste products (e.g., carbon dioxide, dimethyl sulfide, nitrous oxide, and methane) into the water, where those compounds move down the oceanic layers from the surface ocean to the twilight zone through the process of physical mixing. In addition, marine phytoplankton aggregate formations and detritus from the ocean surface drops down into the mesopelagic zone and combines to form the sinking particles of carbon and nutrients, where they settle into the deep ocean and are then decomposed by bacteria. There they are consumed by archaea who produce $CO_2$ through aquatic respiration and will be captured by the migrating zooplankton population in the twilight zone, or they condense their carbon and nutrient storage to create organic carbon. Regardless of the different pathways presented for these sinking particles, all pathways will eventually find their way into sediments and descend to the benthic zone where the carbon and nutrients they carry within them are locked away below the seafloor [70]. Moreover, when the temperature of the ocean's surface water decreases and its salinity increases, the surface becomes much denser than the water beneath, causing it to sink to the deep sea, in turn causing downwelling and deep-water formations which can lock away the $CO_2$ from the surface in the ocean floor. Moreover, ocean upwelling occurs when the surface currents become dislocated from each other, bringing up the deep water to the surface while pushing down the surface water to the deeper layers (i.e., ventilation). This leads to a redistribution of water and with that heat, nutrients, and oxygen within the ocean, fertilizing the surface water and increasing the biological productivity of the surface ocean [71]. With the pathogen/pollutant/nutrient runoff from the coastal land area and the emission of $CO_2$ from the ocean floor sediments during its organic matter decomposition process (i.e., benthic $CO_2$ flux), which produces nitrogen, phosphorus, iron, and silicon, ocean upwelling, downwelling, and ventilation can regulate and better distribute these materials, maximizing the ocean fertilization goals [70]. Because leveraging the ocean's carbon storage capacity is well into the distant future, by artificially accelerating these two processes (i.e., phytoplankton population activities & ocean upwelling/downwelling), the ocean carbon extraction effectiveness and efficiency can be greatly improved.

As ocean fertilization is implemented to accelerate a natural process that extracts $CO_2$ from the atmosphere, it stands to reason that it is a relatively safe practice and technology. However, due to the lack of research on certain aspects of marine carbon and nutrient cycling that has not been fully explored, some unexpected problems may be created from the fertilization of ocean phytoplankton populations and the cumulative effects of ocean up/downwelling. Not only will ocean fertilization aid in the ocean alkalinity enhancement strategy to reduce seawater acidity, but it can also serve to pull more $CO_2$ from the atmosphere in less time, as shown in Table 10, where each cycle would only have 1 week of lasting effects with relatively high annual rate and potential of carbon storage in the ocean. Nevertheless, ocean eutrophication (i.e., "excessive richness of nutrients in a body of water, frequently due to runoff from the land, which causes a dense growth of plant life and death of animal life from lack of oxygen" [72] presents as a disadvantage of ocean fertilization, whereby the nutrient needs of the ocean may be exceeded, causing potential negative side effects. Ocean fertilization on the phytoplankton population has a short-term effect and requires further research to be conducted for confirmation of longer-lasting results.

Calculations and data collection based on the "current technological readiness [and] the time needed to reach full implementation" [73] reflect the technological feasibility of ocean fertilization and is comparably low to other mitigation and adaptation technologies (e.g., reef restoration, renewable energy, vegetation, etc.) Furthermore, the cost-effectiveness of ocean fertilization is relatively lower than most other technologies in the same broad category, resulting in a desired "good" result with lesser economic input. Ultimately, to achieve high levels of carbon extraction without provoking ocean eutrophication and/or other negative effects, governments are recommended to establish restrictions, regulations, and distribution of resources which promote ocean carbon uptake only within tight controls.

Table 10: Ocean Fertilization Data Analysis

| Work Cited: [74] [75] [76] | Lasting effect period (week/cycle) | Global annual $CO_2$ removal potential (GtCO2) | Global ocean $CO_2$ storage (100-year net carbon sequestered) | Surface area needed $(km^2/GtCO_2)^a$ | Total cost for implementation of practice ($/tCO_2$) |
|---|---|---|---|---|---|
| *Ocean Fertilization* | ≈ 1 | $(1 - 2)^b$ | 0.4% - 8.3%: carbon biomass 2% - 44% : carbon exported through iron fertilization | ≈ 1,000 | $(30 - 60)^b$ |

a: Gigaton of $CO_2$
b: "x - y" represents the domain and limitation of the variable from x to y

*Artificial Sand Dunes and Dune Rehabilitation:*

Worldwide, including the 95 nations and states that appear as islands, the total shoreline length stands at 356,000 km and the total coastal area globally, including the land (148.94 million $km^2$) and water (361.132 million $km^2$) portions, amounts to 510.072 million $km^2$ [77]. Using these natural resources, artificial dunes and dune nourishment can be widely distributed and implemented as an adaptation technology in response to oceanic and coastal threats introduced by climate change. The goal of artificial dunes and dune regeneration is similar to the construction of seawalls where both aim to establish a barrier between the sea and the land, protecting the residents and natural habitats from coastal erosion and flooding. The dynamic ability of dunes, whether artificially or naturally assembled, enables this technology to adjust and shift in shape and size as the sea level, ocean currents, wind, and wave climate fluctuates; therefore, allowing the dunes to supply and store sediments to the beach according to their prevailing environment. In total, there are 5 general types of sand dunes: "transverse, linear/longitudinal, star, barchan/crescentic, and parabolic/blowout," [78]. Whether accomplished by artificial construction or by natural formation, dredged sources as well as naturally occurring deposits such as mud and sediment on the coastal regions of the beach can create or restore the dunes. Furthermore, by attaching supplementary defense structures (e.g., fences, planted vegetation) on these dunes, wherein such fences built next to the sea are constructed with natural materials that can easily decompose while vegetation can collect sediments near their location, both are done to promote dune growth, trap sand, and stabilize dune/sand surfaces. Artificial dunes and dune stabilization are not limited to developed beaches alone. Rather, they can be implemented on a variety of land, including "existing beaches, beaches built through nourishment, existing dunes, undeveloped land, undeveloped portions of developed areas [,] areas that are currently fully developed but may be purchased so that dunes can be restored," minimizing the limitations and prior restrictions of sand dune creation and restoration [79].

Different from sea walls, dune nourishment/creation occurs and can be maintained more naturally, leaving less waste and pollutants on the coasts. Additionally, the dunes contribute largely to the maintaining of wide coastal zones on the beaches they reside in. Further, dunes can dissipate wind, wave, and storm energy and present an ablative barrier to coastal erosion where sand from the dunes will be eroded away during different seasons, coming to rest as sediments at the bottom of the coastal regions' waters instead of decreasing landmass of beaches that are being eroded without the protection. Likewise, due to the protection provided by the dunes that shield local inland residents from coastal erosion and flooding, sustainable commercial and other development are more likely to be promoted as a result. This, in turn, benefits local regions economically, just as naturally created dunes can benefit the residents, habitats, and organisms environmentally and ecologically [79].

Although dune regeneration and creation can be implemented with fewer costs and are flexible with many other mitigation and adaptations technologies, some negative side effects of this practice, more specifically with the introduction of new dunes, can still be encountered. Not only will the dunes present as a barrier that protects the residents from oceanic hazards, but it also creates an obstruction that can impede residents from beach access that was once much easier. As well, when implemented in unsuitable regions dunes may create destruction of natural habitats, killing and/or dislocating native species. This result can occur most likely when dune creation and regeneration necessitate that construction areas be zoned off from the public and wildlife residents to maximize the growth process of the dunes. Correspondingly, dunes take up a considerable amount of land area, yet some may not have sufficient protective effects on the land designated for protection as the residents and government may desire. This may commercially and recreationally affect the residents, where land loss may arise as a potential problem, and to the public, where fewer tourist activities may be enabled [79]. Furthermore, even though dunes can dissipate wave energies, some may not be sufficiently robust to stand against strong storms and wave action, thus easily destroyed by such coastal activities. Given that dunes cannot regenerate themselves in a relatively short period, the costly process of reconstruction must then be repeated to maintain the dissipating and erosion process.

Technologically, artificial dune and dune rehabilitation are at a matured and developed level for implementation, where the practice has been adopted for roughly 70-100 years (e.g., the U.S. [1920s], Europe [1950s]) [80]. However, residents and governments are yet to reach a compromise or agreement on the size, type, frequency, and other factors concerning this adaptation technology to avoid conflicts of interest, public opposition, and additional negative effects from both which would otherwise be avoidable. Nevertheless, dunes can serve as an opportunity to educate the public about climate change concerns and the threats posed to Earth's ecosystems, environments, and essentially, their everyday life, physically and mentally preparing residents as well as the public at large for possible future events. As more and more people accept the challenges and dangers climate change brings, dune rehabilitation and creation can be implemented for wider areas, better protecting inland regions as it serves its multi-purpose function.

**Social**

*Raising Public Awareness*

The disastrous effects of climate change on life around the globe are undeniable even as public urgency remains severely below that of which is necessary to bring about effective change. According to a study conducted in 2019, 63% of Americans support climate change policies and believe their necessity is worth the economic cost [81]. This public view, however, is nearly identical to the response given 25 years ago, indicating that there has been little improvement, if any, in public support about global warming's consequences. If, for instance, the decision to mitigate climate change were reduced to a single bill to be passed by Congress, a two-thirds vote by a Congress represented identically with the views of the American public would likely not be achieved and the bill would fail. Although numerous communication campaigns have been established during this period, many are criticized for being inadequate to provoke real action. In addition, although knowledge about climate change may be more advanced with the spread of information via digital media, changing attitudes and behaviors are crucial for enacting actual improvements on the issue. In order to combat the lack of public awareness, Figure 2 provides a detailed and clear layout of the approach to take on such strategy.

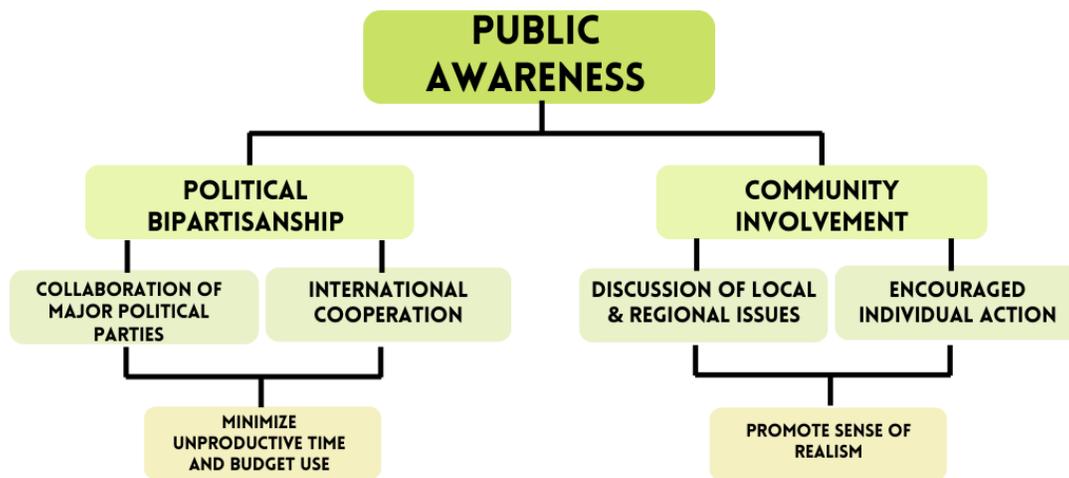

**Figure 2: Raising Public Awareness Model:** The model describes the two branches of major improvement areas (political bipartisanship and community involvement), and the specific actions required to reach the eventual goals of each.

Increasing public awareness can best be arrived at from two different angles: incentivizing political bipartisanship and promoting community involvement with the goals of minimizing unproductive time and budget use and creating a sense of realism, respectively, as shown below in Figure 2.

*Youth Education*

The societal, economic, and welfare impacts from global warming will endure into the indeterminate future, affecting many generations to come unless the threat is successfully and absolutely confronted. Therefore, each generation that inherits the Earth plays a critical role in protecting it and empowering future generations to be knowledgeable about climate change. As such, this should be an important goal in education. Although the scientific community has largely reached a consensus view regarding the importance of youth education to help bring about climate change action, details of such plans are unclear. Most controversy surrounds the topic of the value of the individual's contributions. "The old argument that 'if everyone does their small part, it will make a difference' is, according to some, simply not valid", because individual contributions on a global scale are simply too microscopic [82]. In addition, cooperation of all citizens of the world - or even only most people across the world - is likely impossible unless impactful economic policies are initiated.

Climate education is unfavored by traditional subject-based curriculums. This is apparent in the "compartmentalization" of subjects, such as a split of different areas of science. While climate change touches on a diverse range of topics, traditional education separates these into distinct topics and oftentimes is taught at different grade levels and at various depths, leading to a disadvantageous division of time and energy. In addition, traditional education places heavy emphasis on academic grading and standardized test results, leading to primarily extrinsic motivations for students to learn and participate in mitigating global warming. Once students are independent of these stimuli, they may well no longer feel the need to actively engage as before. Thus, climate education requires the reform of public education, for example, reorganizing academic topics to point out that school subjects are interrelated with each other and are embedded into a complex network of real-world cause and effect. A robust standardized curriculum related specifically to climate education across all states should be considered to ensure future generations are equipped to understand the challenges their generation will face in combating this issue.

Importantly, educators who directly interact with students must be willing to provide climate education and support the cause. Daily interactions between students and teachers can be highly influential on young students' minds and beliefs. One study observed that teachers do not consider "the role of science education to try to solve today's major social, political, economic, technical or scientific problems" [82], which is detrimental to a students' knowledge and views on climate change, especially if such teachers are involved in the child's education from early on. In addition to actively endorsing climate intervention, teachers must be accurately trained in climate education using sources of unbiased data. Misinterpretation, bias, and lack of support must all be eliminated before an educator can satisfy the requirements of climate education training. This consistency must also apply to teachers across different schools, districts, counties, states, as well as teachers of other subjects such as the arts or English, to avoid confusion and doubt. The student will then realize that climate change affects life in general and is not just a remote issue that is discussed only in science class.

Lastly, youth education must be strictly bipartisan and unbiased in every aspect. Teachers, although entitled to their own opinions, should not advocate for their personal beliefs but rather bring different perspectives based on broadly verified facts and sound, logical reasoning, as well as encourage sensible discussion from everyone in the class.

Educators should not fear but rather embrace diverse opinions in the classroom, welcoming these opinions by approaching them with patience and understanding, demonstrating to their students this important component of the climate education discussion. Further, discussions should incorporate both formal and informal elements, for example, technical terminology can be explained in relatively more vernacular language and thus still carry authoritative weight and a sense of reliability while also being more accessible to students' developing level of understanding.

*Domestic Funding*

Domestic funding is closely linked with public awareness and youth education, which if successful, will lead to greater public support and more investments and funding towards mitigation and adaptation strategies. Domestic funding is vital as well for the encouragement of technological innovation and providing financially secured and stable motivation for the continuation of climate change research. One example of a new climate mitigation technology with major potential is the Traveling Wave Technology, which "offers 30 times more efficient use of mined uranium and a factor of five reduction in waste, all based on a once-through fuel cycle without the safety and proliferation concerns of reprocessing used fuel" [83]. Its key characteristic is that it employs depleted uranium — or the "excess" uranium that is not fissile — to generate nuclear energy, thus enabling a significantly more efficient method of obtaining nuclear energy. Similar technologies that are still in the earlier research stages must have sufficient funding to continue development.

Before 2017, the United States had been one of the largest contributors towards financing climate change action; however, this trend was halted with former President Donald Trump. Current President Joe Biden budgets more than $36 billion to combat global warming, including $10 billion for clean energy innovation, $7 billion for NOAA research, $6.5 billion for rural clean energy storage and transmission projects, $4 billion for advancing climate research, $3.6 billion for water infrastructure, $1.7 billion for retrofitting homes and federal buildings, $1.4 billion for environmental justice initiatives, plus another $21 billion on research.

In addition, domestic funding will very likely originate from economic needs for conversion instead of from environmental concerns, unless areas of the United States are damaged or otherwise experience direct and irrefutable consequences from climate change specifically. While environmental complaints are beneficial to reformation, true change will require a certain critical amount of economic and financial momentum in order for politicians and policymakers to become sufficiently incentivized to initiate them.

## 3. Combination

*Optimal Combination:*

After analyzing the solutions above in terms of potential effectiveness, financial feasibility, current readiness, and most importantly, compatibility, we present an optimal combination as a suggestion and reference for governments when making political decisions regarding climate change and corresponding actions, as shown below in Figure 3.

As the combination draws from solutions in various topics, it has a better likelihood of resolving more aspects of the fundamental problem, complementing one another to increase benefits. Incorporating both mitigation (M) and adaptation (A) strategies, this combination merges and emphasizes the benefits acquired both from selected negative emissions technologies (i.e., afforestation, ocean alkalinity enhancement, and bioengineering) and governmental policy solutions (i.e., cap-and-trade, clean energy industry establishment and expansion, and international contract proposals) while constraining the concerns and side-effects generated by each strategy to a minimum through a beneficial cycle of systems.

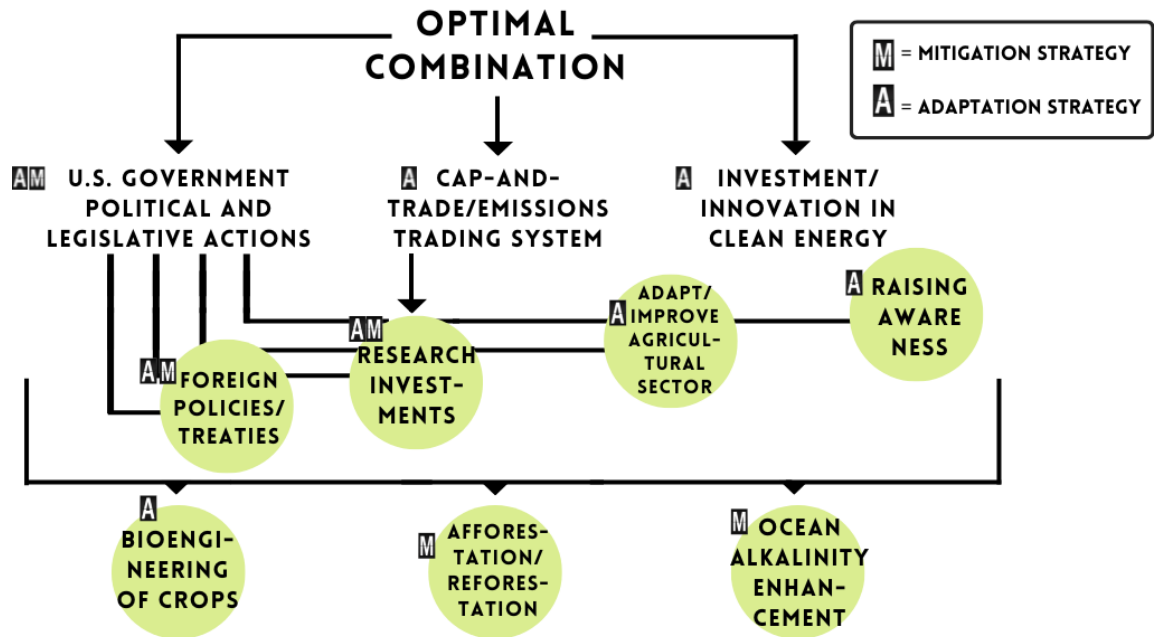

**Figure 3: Optimal Combination Model:** The model labels each proposed solutions as "M" (mitigation strategy), "A" (adaptation strategy), or "AM" (both mitigation and adaptation strategy), and explains the relationships and interconnectedness between each solutions, where the political (U.S. Government Political and Legislative Actions), economic (Cap-and-Trade/Emissions Trading System & Investment/Innovation in Clean Energy), and social strategies (Raising Awareness and Public Education) enables the implementation of technological mitigation (Ocean Alkalinity Enhancement and Afforestation/Reforestation) and adaptation (Bioengineering of Crops) strategy.

To commence, climate intervention will have to come about through economic forces driving countries and corporations to change their former methods and adopt new net carbon free practices that they decide are most efficient and cost-effective. Thus, the need for an economic framework that compels policymakers and others with the potential to bring about change (e.g., corporation leaders) to reduce the need for energy derived from pollution heavy sources is inevitable. From this, a cap-and-trade system is recommended, as described in Figure 3, since compared to the other proposed economic system mentioned in the paper -carbon tax - it presents more flexibility in the marketplace. For instance, minimum government intervention will produce the best price for carbon credits and establishment of this price would be assimilated into the market automatically. On the other hand, a carbon tax would have the opposite effect on businesses as its strict regulatory regime would lead to overall dissatisfaction from the entire industrial sector, making cooperation more challenging and risks politically charged lobbying or price spikes in consumer goods and services. Moreover, cap-and-trade systems bring with them more freedom to consumers by allowing them to shift their purchasing from a given company to its competitors offering relatively lower cost products because they utilize the trading aspect of cap-and-trade. The "Cap and Trade/Emission Trading System" category in Figure 3 emphasizes that the system not only allows the government to better control the carbon emissions of these large corporations, but it can also generate additional revenue that can be implemented in research and development investments, serving as a support for other technology's implementation.

In addition to being economic drivers, direct financial investments– -including government subsidies, donations, private investments, and fundings from NGOs - are just as important. Specifically, research and investment shall be directed into renewable energy, as described under the branch of "Investment/Innovation in Clean Energy" in Figure 3. Heated debates have occurred between supporters of nuclear energy and renewable energy, and this disunity of the scientific and political communities have greatly hindered progress in legitimate action. Therefore, to ensure all aspects of the issue have been addressed, a blend of both clean energy sources is recommended as a greater positive effect can be achieved with emphasis set around a clearly defined goal that involves both rather than unproductive disputes which only serves to squander time and avert attention from the environment to politics. A coexistence of renewable and nuclear energy allows these clean energy forms to be utilized to their maximum potential. For instance, nuclear plants can be established away from residential areas while highly localized renewable sources such as home and office-based solar panel installation will be subsidized to compensate for the expense of nuclear power plant construction and expansion.

Furthermore, the U.S. government itself must ensure that it is maintaining necessary progress in terms of political and legislative actions. Specific suggestions for legislative decisions are listed in Figure 3, under the "U.S. Government Political and Legislative Actions" section. With the goal to increase support towards such efforts, raising awareness is vital and thus so is the need for reform in the education system. All schools and other educational facilities in the U.S. need to incorporate a consistent and robust study of climate change into the curriculum. Detailed precautions of implementing this curriculum can be found above. Second, the U.S. government needs to support adaptation for its agricultural sector. As climate change impacts the country, the agricultural sector faces especially destructive consequences from severe weather and the longer-term implications of a changing climate. Adaptation measures such as improving infrastructure, constructing dams or other forms of crop protection, and subsidies to farms most directly exposed to the impacts should be considered. Third, scientific research targeted towards the sustainability of crops must be conducted to support maximum yields. The protection of the agricultural industry is crucial as it is responsible for maintaining the nation's food supply and occupies a major role in international trade. Extreme events caused by climate change, such as droughts and floods, have historically crumbled certain areas of agriculture. Fourth, technological advancements should be supported. This refers to all types of technology for combating climate change efforts including the development of new, more efficient NETs and improvement of current methods. Fourth, the US government needs to match and aid in efforts by non-governmental organizations already involved in research and investment. The budget would come from either revenue generated by the cap-and-trade system recommended above and/or cuts in funding for less urgent issues. Finally, the government ultimately needs to interact with other countries through foreign policies. For instance, cooperation, discourse, economic pressure, and potentially political pressure are most of the time necessary for America to initiate a chain of desired actions. Drafting well-intentioned treaties, although commendable, ultimately will result in a lack of legitimate action if administered with poor oversight or supervision. Stepping beyond the stage of only discourse and into concrete actions is now needed to move forward on improved efforts for cooperation and results on a global scale. Consequently, governments are encouraged to allocate a considerable portion of their total budgets for climate change mitigation and adaptation technology research and development purposes (e.g., ~5% of total spending), this redirected from overall growth of revenue and otherwise continuously escalating military budgets. With better utilization of capital resources, countries increase the likelihood to fund critical technologies to combat climate change, resulting in more realistic goals and achievements that can reduce climate change damage and threats far more effectively.

As shown in Figure 3, the above actions serve as a synergistic economic and political support system for the implementation of technological mitigation and adaptation technologies. Thus, the technologies most suitable to be in the optimal combination for maximizing positive impact in combating climate change are bioengineering of crops, afforestation/reforestation, and ocean alkalinity enhancement. In the case where no action is conducted in response to climate change, the global environment will be largely degrading worldwide, punctuated by declining condition of soils (e.g., salinification and desertification) and habitats (e.g., increase of temperature, water shortage, and loss of habitats in general) causing and exacerbating many problems within individual nations and society in general (e.g., food insecurity, economic inequalities, etc.) In response to this concern, bioengineering of crops serves to upgrade their adaptability and creates more crops with desired characteristic for their growing conditions. Subsequently, this allows for more crops to be produced within a more compact land area which reduces food insecurities and excessive water usages by keeping pace with the demands for food production and storage to feed rapidly growing populations. In areas that previously proved inefficient to support large quantities of agricultural plants to be grown and harvested, crop bioengineering allows local and regional farmers to select appropriate crops that are genetically modified to withstand the prevailing harsh environment, therefore making use of many wastelands or empty spaces that would otherwise not be usable. By maintaining a steady production of food, the impact of climate change on people's lives will be substantially diminished, enabling society a longer period to counter climate change while minimizing serious consequences such as famine and conflicts over resources.

**4.0 Summary**

By implementing technologies through acceleration of natural mitigation processes found in forests (i.e., afforestation/reforestation) and oceans (i.e., alkalinity enhancement), the negative environmental effects are reduced to a manageable, controlled rate with benefits that are much more predictable. Given that Earth's soil and ocean carbon storage capacity well exceeds many other methods and the processes required demand much less economic investment than many other more technologically challenged approaches, these can be conducted to scale over a long period of time to mitigate the desired amount of $CO_2$ from the atmosphere with little concern of reaching carbon storage capacity or economic limitations, as shown below in Table 11. With minimized interference versus other technologies, afforestation and ocean alkalinity enhancement can be conducted over large portions of the globe without incurring serious social, economic, or environmental disputes. Moreover, if the technologies studied and listed in Table 11 are implemented in a manner which maximizes benefit, not only will these technologies prove advantageous to the environment, but can also benefit local habitats by restoring many that have been lost due to degrading natural structures while improving residents' living conditions socially (e.g., lessen unemployment rate, reduce food insecurity, etc.) and economically (e.g., boost of food production and trade). With all three steps combined, countries can work together within their states and provinces to maximize the beneficial effects of their applied technologies, Mitigation techniques included in Table 11 can be paired with adaptation technologies examined above to slow the rise in, and eventually lower, atmospheric $CO_2$ efficiently and effectively with realistic and adequate economic support and investments from governments and private industry.

Table 11: Mitigation Technologies Data Analysis

| Measurable mitigation strategies | Cost per Gt of $CO_2$ mitigated ($) | Total $CO_2$ mitigation capacity by 2030 (Gt) | Technological readiness for large-scale implementation |
|---|---|---|---|
| *CCUS* | $100 billion - $1 trillion | N/A | Yes |
| *BECCS* | $20 - $100 billion | 0.5 - 5 (2050) | Yes |
| *Afforestation/Reforestation* | $0 - $50 / $104 billion | 3.6 (2050) | Yes |
| *Geologic Reservoir Sequestration* | $7 - $30 billion | 62.5 (2050) | No |
| *Soil Carbon Sequestration* | $0 - $100 billion | 1 - 5 (2050) | Yes |
| *Ocean Fertilization* | $18 - $60 billion | 3.2 - 9.4 | No |
| *Ocean Alkalinity Enhancement* | $55 - $160 billion | 2.5 - 10 | No |
| *Carbon Tax ($40)* | +$26 billion net revenue | 20 | Yes |
| *Cap and Trade ($40)* | +$7.9 billion net revenue | 38 | Yes |

**Note:** Summarizes the mitigation solutions to provide a comparison of their effectiveness through their cost per Gt of $CO_2$ mitigated, total $CO_2$ mitigation capacity by 2030, and technological readiness for nation-wide implementation.

## 5. Conclusion

The content of this paper could essentially be separated into two parts. The first consists of a detailed analysis and breakdown of almost twenty-five different climate change combating solutions, ranging from a cap-and-trade system to stratospheric aerosol injection. These proposals include both mitigation solutions, referring to those that directly decrease greenhouse gas emissions per year or total quantity in the atmosphere, and adaptation solutions, which are those that prepare vulnerable communities to better face the consequences of climate change. Arranged into seven categories, the approaches listed are classified as energy (i.e., nuclear & renewable), economic and political (i.e., carbon tax, cap-and-trade, research & investment, government subsidies, and direct/indirect aid to other countries), agricultural and agroforestry (i.e., afforestation, reforestation, bioenergy carbon capture and storage [BECCS], bioengineering [BE] of crops/genetically modified organisms [GMOs], and irrigation systems), atmospheric and astronomical (i.e., carbon capture, utilization & storage, stratospheric aerosol injection, marine sky brightening, and space-based mirrors), geological (i.e., geologic reservoir sequestration and soil carbon sequestration), coastal and oceanic (i.e., ocean alkalinity enhancement, ocean fertilization [OF], and artificial sand dunes and dune rehabilitation), or social (i.e., raising public awareness, youth education, and domestic funding) applications. The analysis of each solution includes a detailed description of its functions, advantages and disadvantages, numeric data, and/or any historic implementations.

Since it is impractical for governments to attempt to utilize all twenty-three solutions at once, only a selected few should be chosen for implementation. The second part of the paper provides the most optimal combination, considering the perspective of the U.S. government at the present time, to achieve maximum potential positive outcomes. It is important to note that combining certain solutions together can provide unique benefits that would not exist if any one of them were to be implemented individually. In this section, the paper explains the reasoning for the selection of every solution in the optimized combination and why those would outperform other solutions in their respective categories, along with how these chosen solution components can enhance the effectiveness of other component solutions contained in the optimized group. The final combination includes the implementation of a cap-and-trade system, an energy industry reformation plan, recommended actions to be taken by the U.S. government (i.e., education, research, foreign aid), negative emissions mitigation solutions (i.e., afforestation, ocean alkalinity enhancement), and an adaption solution by way of cautious bioengineering.

As the effects of climate change are nearing irreversibility, we sincerely and strongly suggest governments of the Earth to take into careful consideration the proposal and unite together to combat this serious challenge that all of humanity faces.

**Acknowledgement:** This study was supported by the Jet Propulsion Laboratory, California Institute of Technology, under contract with NASA.